\RequirePackage{fix-cm}

\documentclass[twocolumn]{svjour3}          

\smartqed  
\usepackage{graphicx}
\usepackage{caption}

\usepackage{xcolor}

 \journalname{Exp Astron}

\usepackage[colorlinks=true,breaklinks=true,linkcolor=magenta,citecolor=blue,urlcolor=green]{hyperref}

\newcommand\cts{counts~s$^{-1}$}

\newcommand\apj{{Astrophys.\ J.\ }}%
\newcommand\apjl{{Astrophys.\ J.\ }}%
\newcommand\aap{{Astron. Astrophys.\ }}%
\newcommand\nat{{Nature\ }}%
\newcommand\procspie{{Proc.~SPIE\ }}%
\newcommand\pasj{{PASJ\ }}%

\begin{document}

\title{PolarLight: a CubeSat X-ray Polarimeter based on the Gas Pixel Detector 
}

\titlerunning{PolarLight}        

\author{
Hua Feng$^{1}$ \and 
Weichun Jiang$^{2}$ \and 
Massimo Minuti$^{3}$ \and 
Qiong Wu$^{1}$ \and 
Aera Jung$^{1}$ \and 
Dongxin Yang$^{1}$ \and 
Saverio Citraro$^{3}$ \and 
Hikmat Nasimi$^{3}$ \and 
Jiandong Yu$^{4}$ \and 
Ge Jin$^{5}$ \and 
Jiahui Huang$^{1}$ \and 
Ming Zeng$^{1}$ 
Peng An$^{4}$ \and 
Luca Baldini$^{3}$ \and 
Ronaldo Bellazzini$^{3}$ \and
Alessandro Brez$^{3}$ \and
Luca Latronico$^{3}$ \and
Carmelo Sgr\`{o}$^{3}$ \and
Gloria Spandre$^{3}$ \and
Michele Pinchera$^{3}$ \and
Fabio Muleri$^{6}$ \and
Paolo Soffitta$^{6}$ \and
Enrico Costa$^{6}$
}

\institute{
H.~Feng \at \email{hfeng@tsinghua.edu.cn} \and
$^{1}$Department of Engineering Physics and Center for Astrophysics, Tsinghua University, Beijing 100084, China \and
$^{2}$Key Laboratory for Particle Astrophysics, Institute of High Energy Physics, Chinese Academy of Sciences, Beijing 100049 \and
$^{3}$INFN-Pisa, Largo B. Pontecorvo 3, 56127 Pisa, Italy \and
$^{4}$School of Electronic and Information Engineering,  Ningbo University of Technology, Ningbo, Zhejiang 315211, China \and
$^{5}$North Night Vision Technology Co., Ltd., Nanjing 211106, China \and
$^{6}$IAPS/INAF, Via Fosso del Cavaliere 100, 00133 Rome, Italy
}

\authorrunning{Feng et al.} 

\date{Received: date / Accepted: date}

\maketitle

\begin{abstract}

The gas pixel detector (GPD) is designed and developed for high-sensitivity astronomical X-ray polarimetry, which is a new window about to open in a few years. Due to the small mass, low power, and compact geometry of the GPD, we propose a CubeSat mission Polarimeter Light (PolarLight) to demonstrate and test the technology directly in space. There is no optics but a collimator to constrain the field of view to 2.3 degrees. Filled with pure dimethyl ether (DME) at 0.8 atm and sealed by a beryllium window of 100~$\mu$m thick, with a sensitive area of about 1.4~mm by 1.4~mm, PolarLight allows us to observe the brightest X-ray sources on the sky, with a count rate of, e.g., $\sim$0.2~\cts\ from the Crab nebula.  The PolarLight is 1U in size and mounted in a 6U CubeSat, which was launched into a low earth Sun-synchronous orbit on October 29, 2018, and is currently under test. More launches with improved designs are planned in 2019. These tests will help increase the technology readiness for future missions such as the enhanced X-ray Timing and Polarimetry (eXTP), better understand the orbital background, and may help constrain the physics with observations of the brightest objects. 

\end{abstract}

\section{Introduction}
\label{sec:intro}

X-ray polarimetry in the keV band has drawn great interests in astrophysics~\cite{Kallman2004} but has remained unexplored since 1970s \cite{Weisskopf1976,Weisskopf1978,Weisskopf1978a}.  Along with the breakthrough in detection technology that makes 2D electron tracking possible \cite{Costa2001,Bellazzini2007b,Bellazzini2010,Bellazzini2013}, high-sensitivity X-ray polarimetry becomes possible and a number of space missions dedicated to or capable of X-ray polarimetry have been proposed \cite{Soffitta2013,Jahoda2014,Weisskopf2016,Zhang2016}. Active missions or mission concepts include the Imaging X-ray Polarimetry Explorer (IXPE) \cite{Weisskopf2016}, which was selected by NASA and is scheduled to launch around 2021, and the enhanced X-ray Timing and Polarimetry (eXTP) \cite{Zhang2016}, which is a Chinese-European collaboration aiming for large-area X-ray polarimetry jointly with timing and spectroscopy.  Both IXPE and eXTP have adopted the gas pixel detector (GPD) as the focal plane detector. Thus, a flight test of the detector is indeed needed.

\begin{figure}
\centering
\includegraphics[width=\columnwidth]{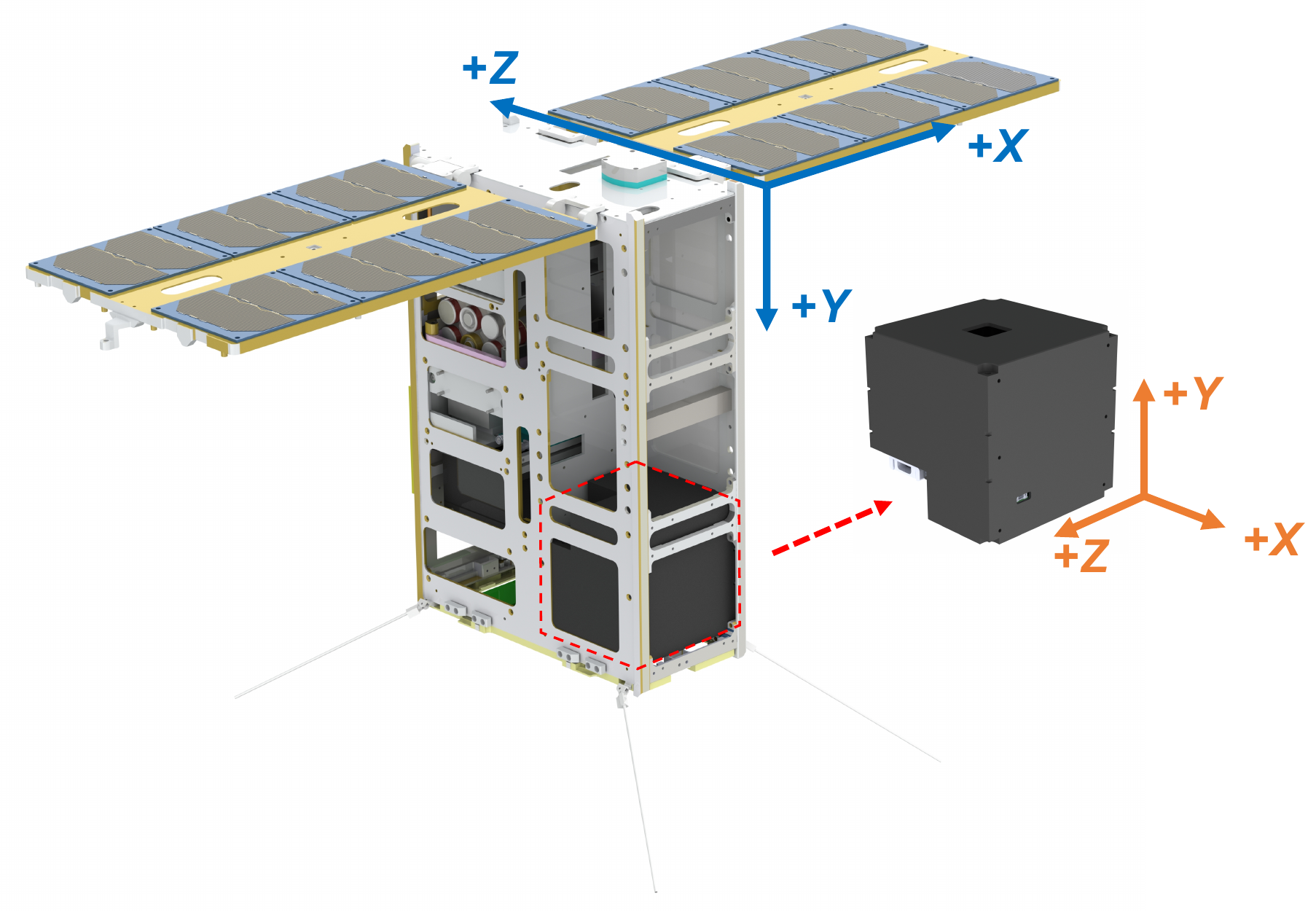}
\caption{Schematic drawing of the CubeSat. The PolarLight is mounted at the side away from the solar panel, with the window pointing toward the $+Y$ direction.}
\label{fig:sat}
\end{figure}

The GPD polarimeter has the advantage of compactness and low mass, and can be operated at room-temperature with a total power of about 2~W. These indicate that it can easily fit into a CubeSat spacecraft. Thus, based on the GPD tested in the lab \cite{Bellazzini2007b,Li2015}, we modified the design and interface to be compatible with a 6U CubeSat developed by Spacety Co.\ Ltd, trying to test the technique in orbit (see Figure~\ref{fig:sat}).  The detector occupies a standard unit of the CubeSat and is named Polarimeter Light (PolarLight), but it is not the only payload of the CubeSat. The satellite was successfully launched into a nearly circular Sun-synchronous orbit on October 29, 2018, with an altitude of about 520~km and an orbital period of 95 minutes. The CubeSat is equipped with a star tracker and three reaction wheels, enabling a pointing accuracy of about 0.1$^\circ$ to a celestial object. On November 16, the PolarLight was powered on briefly and tested with charge injection (injection of charge to the preamplifier of a pixel), indicating that the electronic part is working. On December 18, the high voltage was applied for the first time and tracks triggered by X-rays and charged particles were seen. After roughly three months since launch, tests for the data transfer and attitude control were successfully done with the CubeSat.  At the time of writing, the PolarLight is able to point at a celestial target and start observations. Here, we describe the structural and electronic design of PolarLight and the ground calibration results. The in-orbit tests will be reported in follow-up papers. 

\section{Structure and components}
\label{sec:struct}

\begin{figure*}
\centering
\includegraphics[width=0.8\textwidth]{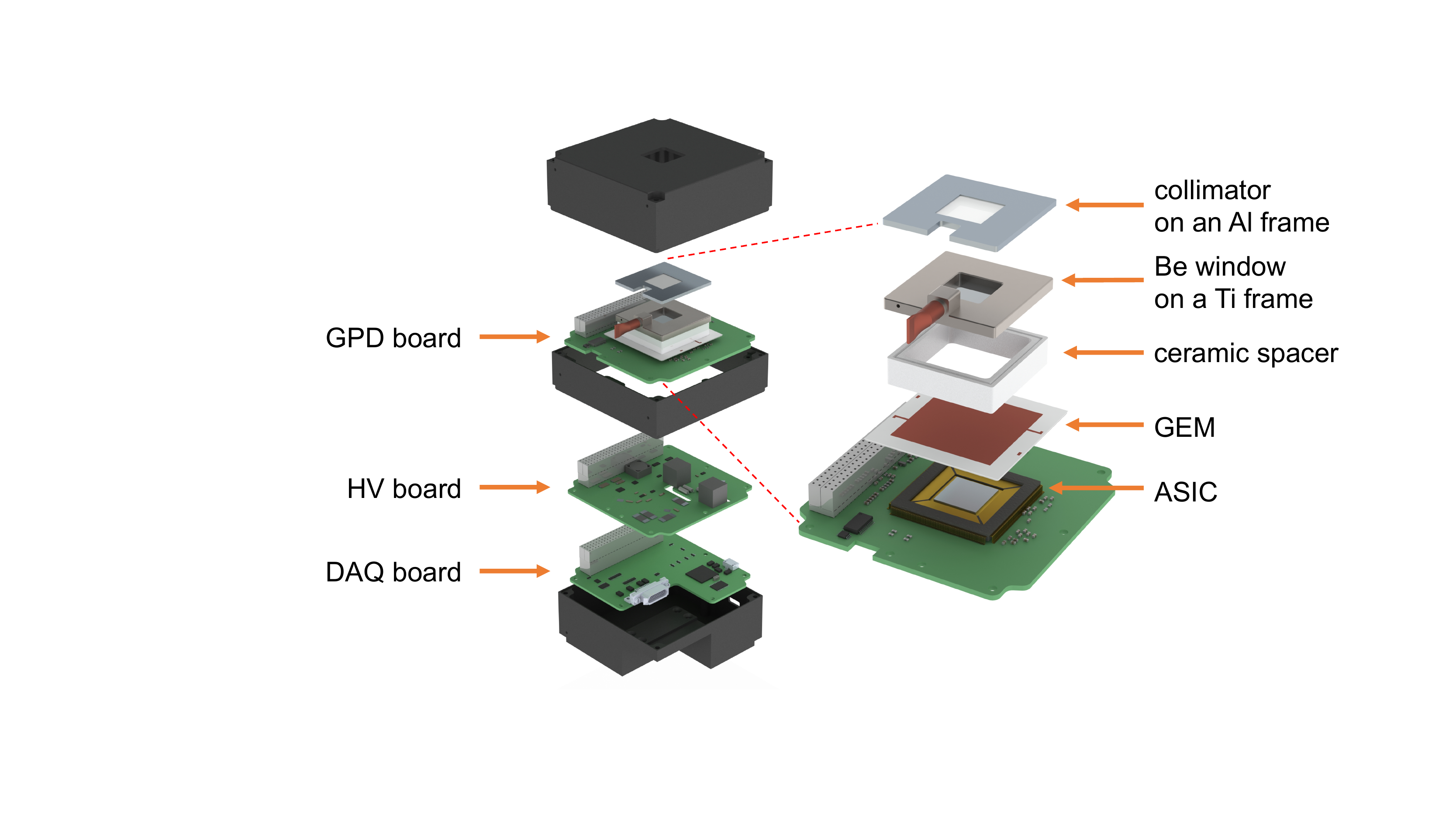}
\caption{Structure of the PolarLight and the GPD.}
\label{fig:polarlight}
\end{figure*}

PolarLight contains three printed circuit boards (PCBs) inside an aluminum case. From top to bottom, the three PCBs are respectively to host the GPD, the high voltage (HV) power supply and dividing circuits, and the data acquisition (DAQ) system. The structure of the PolarLight is shown in Figure~\ref{fig:polarlight}. Here, we elaborate the detailed specifications of some key components.

\subsection{Detector}

The GPD for PolarLight is based on the design introduced by the INFN-Pisa group \cite{Bellazzini2007b}, and is similar to the one reported in Li et al.\ \cite{Li2015}. It is a 2D gas proportional counter with pixel readout to measure the track image of photoelectrons emitted following the absorption of X-rays. A valid event consists of following processes. An incident X-ray goes though one of the holes of the collimator, penetrate the beryllium window, and is absorbed by the working gas of the GPD. After absorption, a photoelectron is emitted and starts to ionize the gas molecules into ions and primary electrons. The primary electrons drift toward the anode under a paralleled electric field. When the electrons go through the holes of the gas electron multiplier (GEM), where the electric field is strong, avalanche happens and secondary electrons are created. The electrons are enhanced  in number by a factor of a few hundred to a few thousand (the gain factor). Then, the secondary electrons will drift along the field lines. Some end at the bottom electrode of the GEM, while the others can go all the way to the anode, which is the ASIC chip. The induced charge on the pixels will trigger the electronics and be integrated, amplified, and filtered by the front-end electronics inside the chip. This is the whole chain how an event is detected. 
 
\begin{figure}
\centering
\includegraphics[width=0.5\columnwidth]{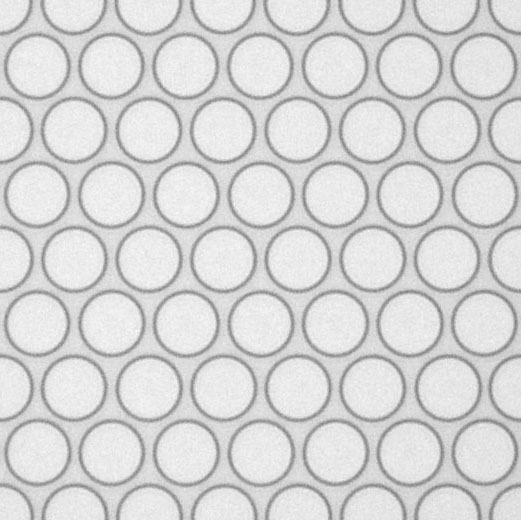}\\
\includegraphics[width=0.6\columnwidth]{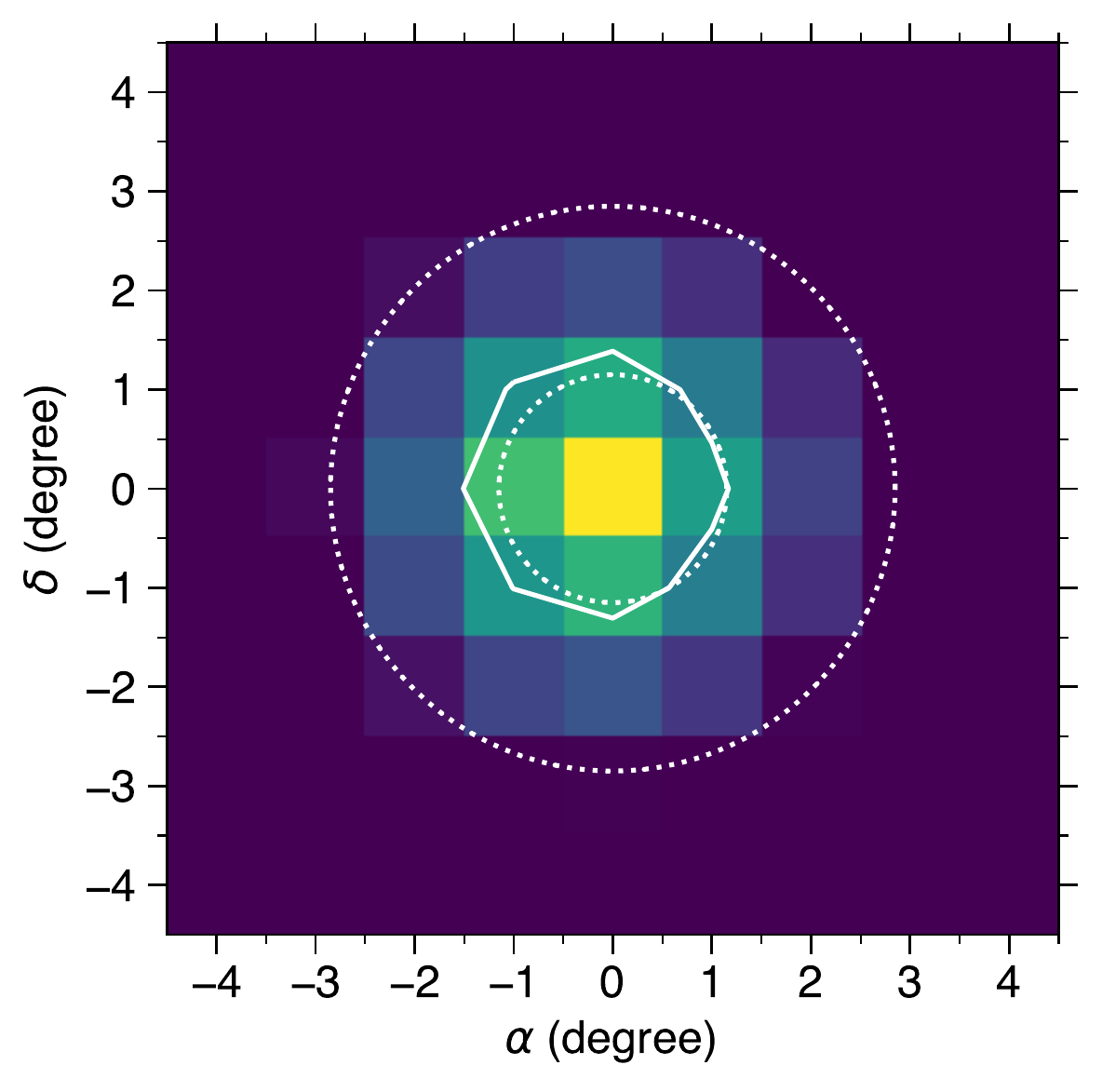}
\caption{\textbf{Top}: microscopic image of the collimator. The collimator is 1.66~mm thick.  The cylindrical apertures have a diameter of 83~$\mu$m and an open fraction of 71\%. The FOV has a FWHM of $2.3^\circ$ and a FWZR of $5.7^\circ$. \textbf{Bottom}: measured angular response of the collimator using an X-ray beam. The solid line is the data contour at half of the maximum response. The dotted lines represents the designed FWHM and FWZR, respectively.}
\label{fig:collimator}
\end{figure}

\begin{figure}
\centering
\includegraphics[width=0.7\columnwidth]{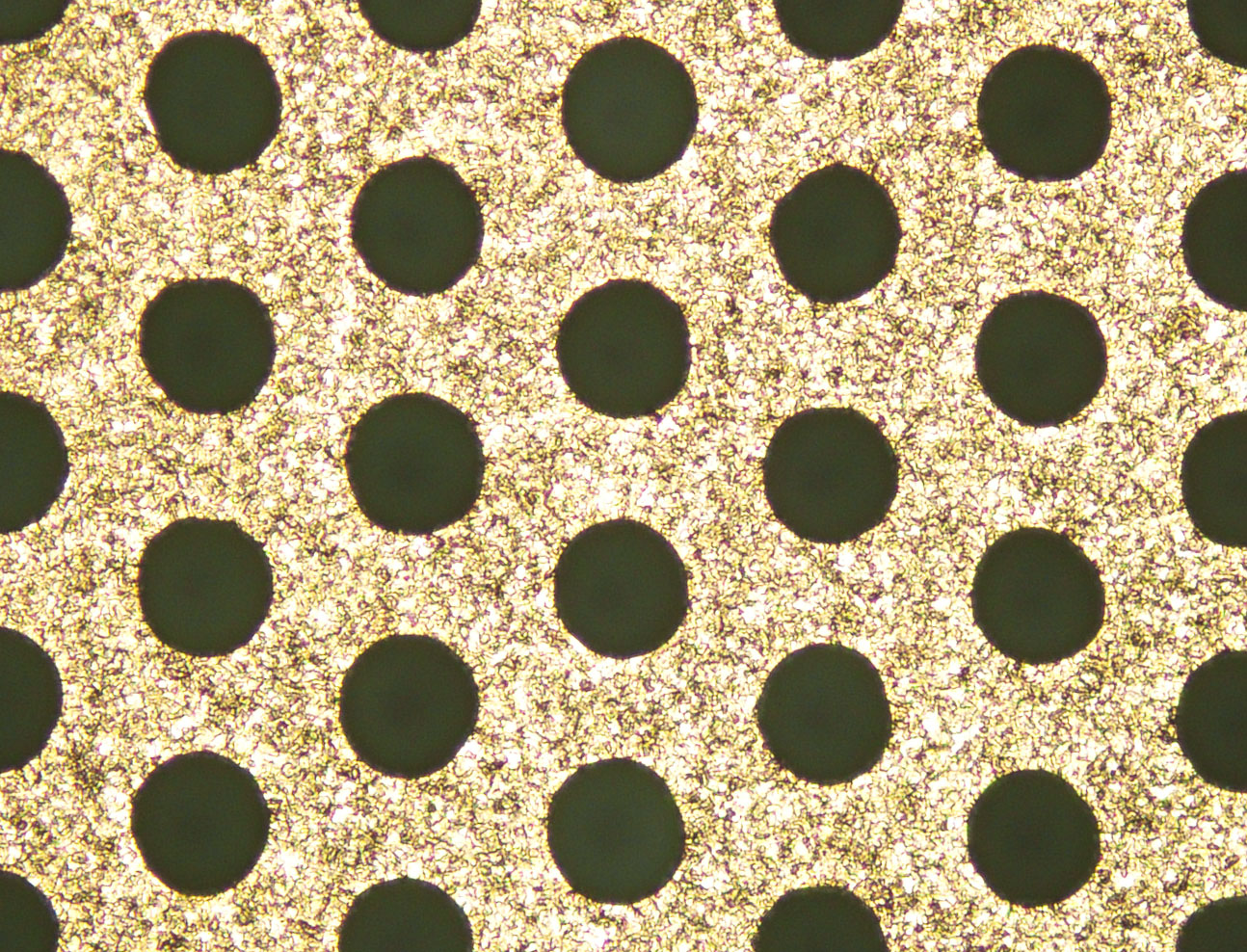}
\caption{Microscopic image of the GEM foil, which is 100 $\mu$m thick. The holes have a diameter of 50 $\mu$m and a pitch of 100 $\mu$m.}
\label{fig:gem}
\end{figure}

\begin{figure}
\centering
\includegraphics[width=0.8\columnwidth]{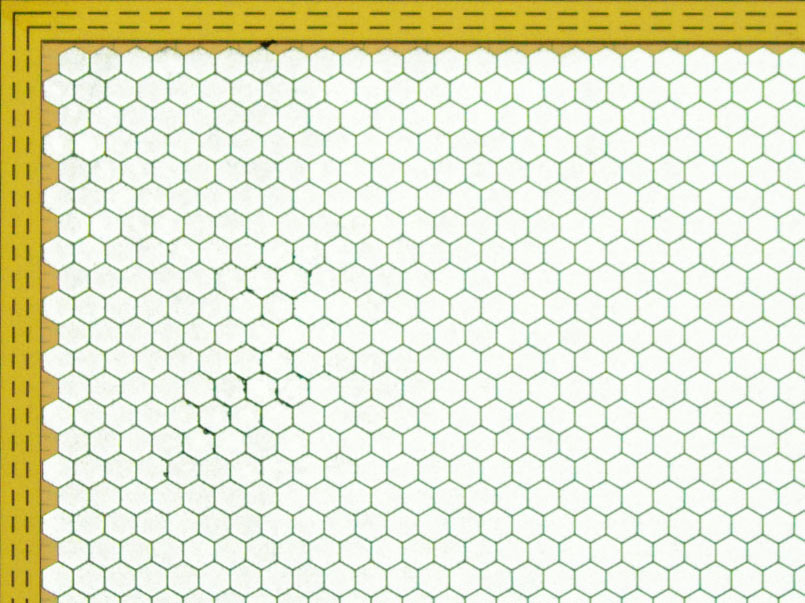}
\caption{Microscopic image of the ASIC chip around the corner.  The hexagonal pixels have a pitch of 50 $\mu$m.}
\label{fig:asic}
\end{figure}
 
\begin{figure}
\centering
\includegraphics[width=\columnwidth]{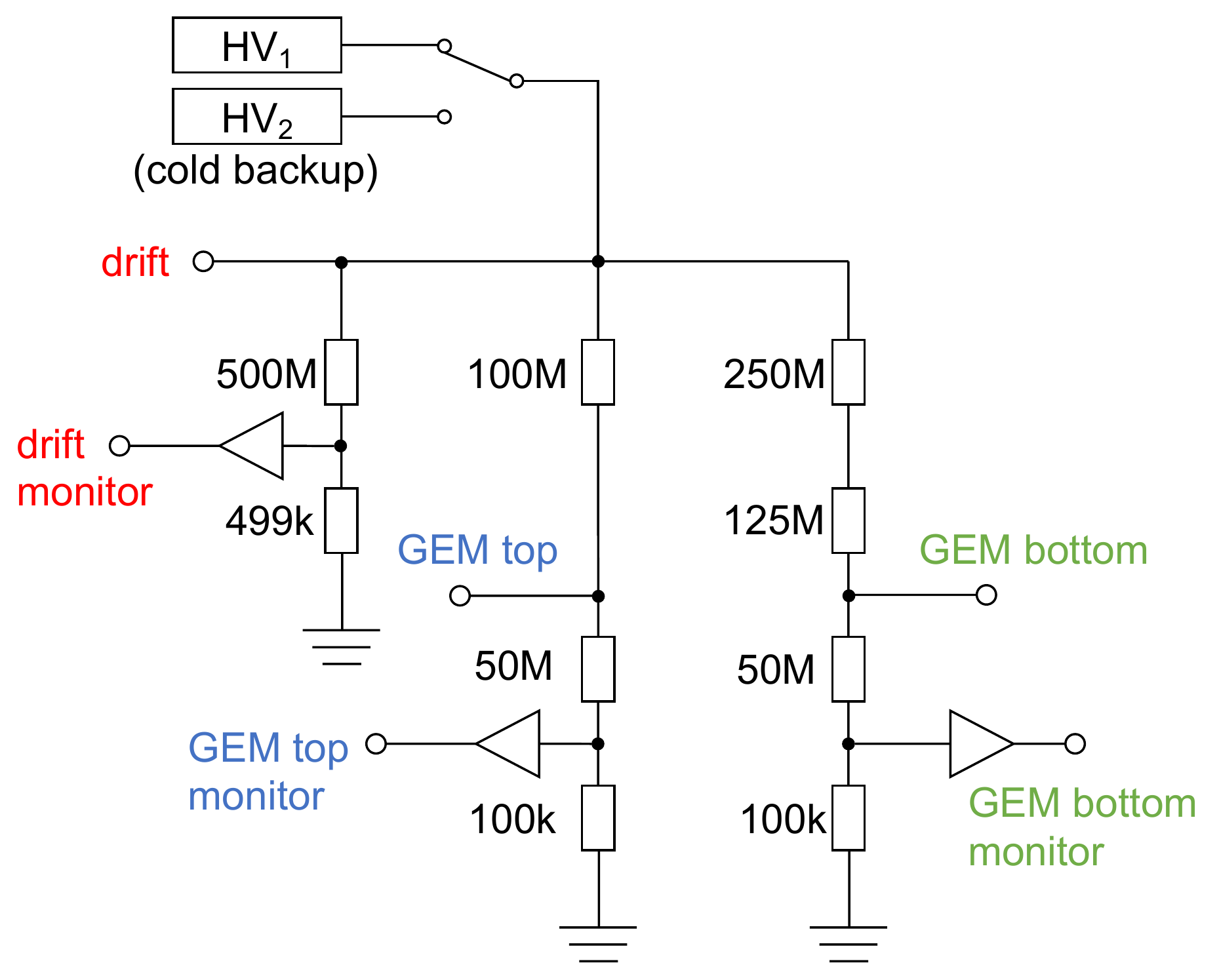}
\caption{High voltage circuits diagram. There are two HV modules for a cold backup.}
\label{fig:hv}
\end{figure}

\begin{figure*}
\centering
\includegraphics[width=0.8\textwidth]{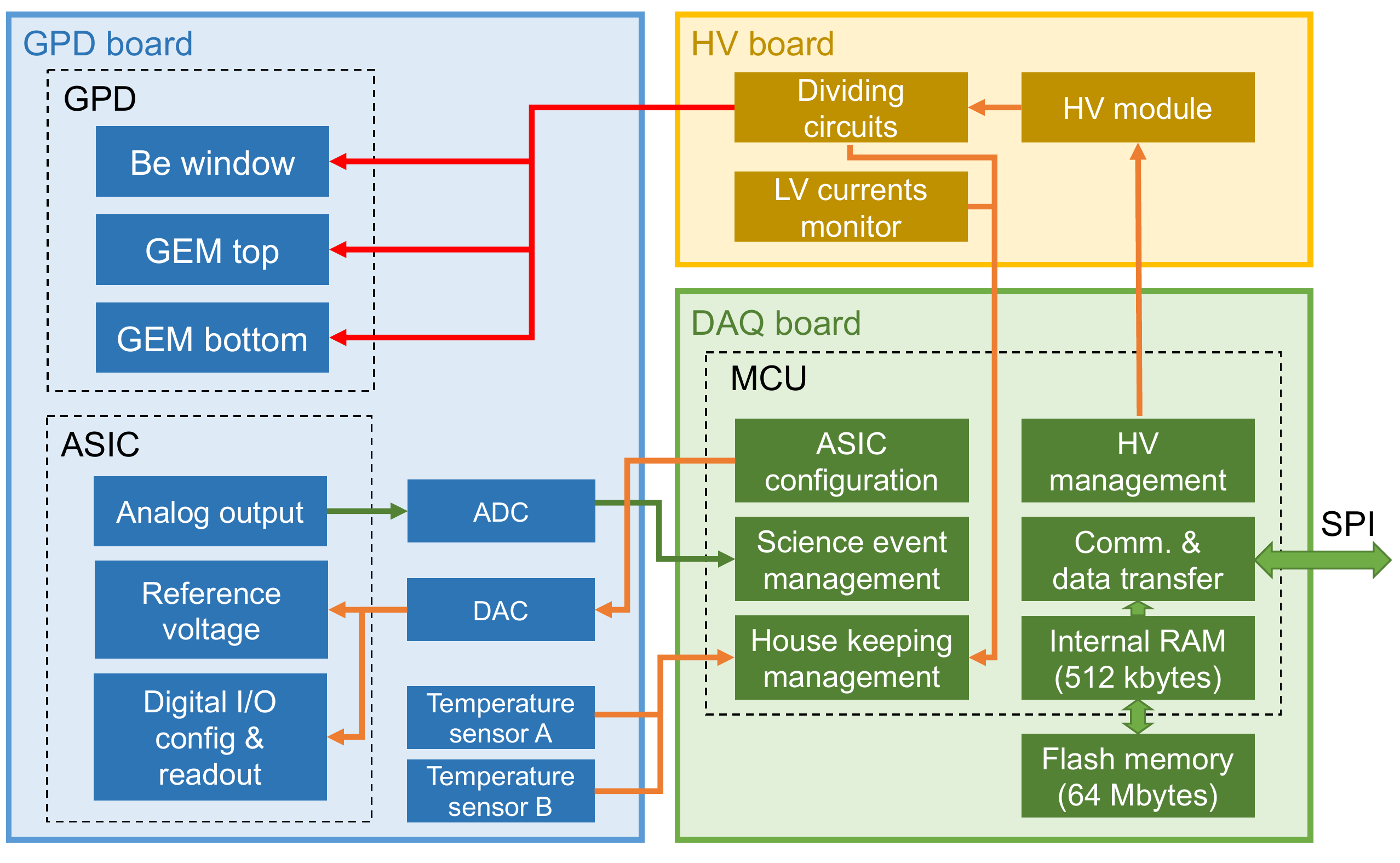}
\caption{Diagram for the back-end electronics of PolarLight.}
\label{fig:electronics}
\end{figure*}
 
Import components of the GPD include the follows. 
\renewcommand{\labelitemi}{$\bullet$}
\begin{itemize}

  \item \textbf{Collimator} --- The collimator is a capillary plate made of lead glass (with $\sim$38\% lead oxide), manufactured by North Night Vision Technology Co.\ Ltd.  It is 1.66~mm thick and contains cylindrical micropores with a diameter of 83~$\mu$m and an open fraction of 71\%. The field of view (FOV) has a  full width at half maximum (FWHM) of $2.3^\circ$ and a full width at zero response (FWZR) of $5.7^\circ$. A microscopic view of the collimator is displayed in Figure~\ref{fig:collimator}.  Using an X-ray tube (Oxford Instruments Apogee 5500 series with a focal spot size of 35 $\mu$m) placed at $\sim$2~m away, we measured the angular response of the collimator and found a FWHM consistent with expected. 

  \item \textbf{Window}  --- The GPD is sealed by a beryllium window of 100 $\mu$m thick. The window is glued and electrically contacted to the titanium frame. When the GPD is in operation, the beryllium window and titanium frame are supplied with a HV of about $-$3~kV, so that the drift field is set at around 2~kV/cm with which the energy resolution and modulation factor is optimized \cite{Li2015}. We note that the whole CubeSat is wrapped by a thermal coat, which was reduced to a single layer right above the window of the GPD. It is a mylar foil of roughly 6 $\mu$m thick coated with aluminum and will further absorb soft X-rays. The Aluminum coating has an unknown thickness, but tests with X-ray sources suggest that the absorption is dominated by the mylar foil.
  
  \item \textbf{GEM}  --- The GEM offers signal amplification by multiplying the number of primary electrons by a factor of a few hundred to a few thousand when they pass through the GEM holes. The GEM foil is a 100 $\mu$m thick liquid crystal polymer (LCP) coated with 5 $\mu$m copper on both sides \cite{Tamagawa2009}, manufactured by SciEnergy Inc., Japan, see Figure~\ref{fig:gem}. The foil consists of laser etched micro-holes with a diameter of 50 $\mu$m and a pitch of 100 $\mu$m in a hexagonal pattern. The operating high voltage ranges from 600--700~V across the top and bottom electrodes, which determines the effective gain in an exponential law. 
  
  \item \textbf{ASIC}  --- The ASIC chip is used for collecting and processing the multiplied charge signals. Figure~\ref{fig:asic} shows the top metal layer of the chip, which is pixelated to hexagonal pixels. Each pixel is connected to a full electronic chain (preamplifier, shaping amplifier, sample and hold, and multiplexer) built immediately below it. The noise is around 50 e$^-$ rms per pixel. The shaping time is 3--10 $\mu$s and externally adjustable. More details about the ASIC can be found in Bellazzini et al.\ \cite{Bellazzini2006b}.
  
  \item \textbf{Gas}  --- The detection gas sealed in the GPD is pure ($>$99.999\%) dimethyl-ether (DME, CH3-O-CH3) at a pressure of 0.8~atm. The choice of DME is because it has a small transversal diffusion coefficient and a relatively high gain, such that the track information will not be diluted after the drift of electrons.
  
\end{itemize}

\subsection{High voltage}

The HV is provided by a compact module UMHV0540N manufactured by HVM Technology, Inc. It has a cubic geometry with a length of 0.5 inch on each side, and is powered by a low voltage power supply of 5~V. With a programming pin, the HV output is adjustable from 0 to $-$4~kV. The output is split into three independent channels, respectively, to power the the drift plate and the two electrodes of the GEM. The dividing circuit is shown in Figure~\ref{fig:hv}. The three HVs are monitored with analog digital converters (ADCs). Two independent HV modules are mounted on the same board, with one of them in use and the other as a cold backup. Wires are used to connect the HVs from the HV board to the GDP board. 

 \subsection{MCU and BEE}

The whole system is controlled and managed by a microcontroller (MCU) TMP570LC4357 manufactured by Texas Instruments. A block diagram of the backend electronics (BEE) is shown in Figure~\ref{fig:electronics}.  At start up, the MCU will configure the ASIC chip. The amplified science signals from the ASIC are digitized by a 12-bit ADC on the GPD board. The data are packed and saved in an external flash memory of 64~Mbytes on the DAQ board.  The DAQ board communicates with the payload computer using an SPI interface, for both commands and data transfer. The absolute timing signals are offered by the global positioning system (GPS) on the CubeSat. The UTC time is broadcasted via the SPI, and the pulse per second (PPS) signal is connected directly from the GPS to the DAQ board. The MCU has an internal timer configured to run at a frequency of 100~MHz. The synchronization between the internal timer, PPS, and UTC is managed by the MCU. 

\subsection{Mass and power }

The total mass of PolarLight is about $\sim$580~g and the total power consumption is about $\sim$2.2~W during normal operation. The detailed mass budget and power consumptions can be found in Table~\ref{tab:mass} and \ref{tab:power}, respectively. 

\begin{table}
\caption{Mass budget of PolarLight}
\label{tab:mass}
\centering
\begin{tabular}{ll}
\hline\noalign{\smallskip}
Component & Mass (g) \\
\noalign{\smallskip}\hline\noalign{\smallskip}
GPD board & 121 \\
HV board & 60 \\
DAQ board & 59 \\
Aluminum case & 322 \\
Accessaries (screws/wires/etc.) & 19 \\
Total & 581 \\
\noalign{\smallskip}\hline
\end{tabular}
\end{table}

\begin{table*}
\caption{Power consumptions of PolarLight}
\label{tab:power}
\centering
\begin{tabular}{ccccc}
\hline\noalign{\smallskip}
HV & Trigger & 5V power (W) & 3.3V power (W) & Total power (W) \\
\noalign{\smallskip}\hline\noalign{\smallskip}
off & off & 1.45 & 0.78 & 2.23 \\
0 V & off & 1.48 & 0.78 & 2.26 \\
0 V & charge injection & 1.46 & 0.52 & 1.98 \\
$\sim$3000 V & off & 1.70 & 0.51 & 2.21 \\
$\sim$3000 V & on & 1.69 & 0.52 & 2.21 \\
\noalign{\smallskip}\hline
\end{tabular}
\end{table*}

\subsection{Data structure and telemetry}

The science data are saved in the event mode. Once there is a trigger, the ASIC will determine a region of interest (ROI) surrounding the triggered pixel. The pixels inside the ROI will be read out and digitized twice, with the first time for signal measurement and the second time for pedestal subtraction. Then the MCU will jump into an interrupt to process and save the event into the memory, along with the precise timing information. As the ROI size may vary, the event block size is not fixed. A description of the science event data structure is listed in Table~\ref{tab:scidata}. 

The house keeping (HK) data are saved every 30~s during X-ray measurements, or every 60~s during charge injection. Each HK package has a fixed size of 434 bytes, containing information about the time, HV, temperatures, data rate, currents from the low voltage powers,  disabled pixels, history commands, and other engineering status. However, the HK saving procedure is set at a lower priority. If the X-ray count rate is high, the time interval for HK saving may be slightly longer. 

The cosmic X-ray background (CXB) for PolarLight is negligible. The particle induced background (internal background) may be dominant. The brightest persistent X-ray source, Scorpius X-1, will result in about 3.5~\cts\ in the detector. The typical number of pixels for events triggered by 4~keV X-rays is about 700. Thus, the total data rate is at least 20~Mbytes per hour if we point the detector at Scorpius X-1.  The CubeSat is operated by Spacety. The commands, telemetry, and small data packages can be transferred through the UHF channel. There are two UHF ground stations for Spacety with 4--6 times of overflight every day. The full data will be transferred to the ground station via the X band, with a chance expected roughly once a week.

\begin{table}
\caption{Science data structure of PolarLight.}
\label{tab:scidata}
\centering
\begin{tabular}{ll}
\hline\noalign{\smallskip}
 & Bytes \\
\noalign{\smallskip}\hline\noalign{\smallskip}
Header & 4 \\
Time & 24 \\
ROI & 8 \\
Image & $n_{\rm pixel} \times 2$ \\
Cyclic redundancy check (CRC) & 2 \\
Tail & 4  \\
Total & $n_{\rm pixel} \times 2$ + 43 \\
\noalign{\smallskip}\hline
\end{tabular}
\end{table}

\subsection{Space qualification tests}

\begin{table*}
\caption{Qualification tests for space environment.}
\label{tab:environ}
\centering
\begin{tabular}{lll}
\hline\noalign{\smallskip}
Test & Date & Conditions \\
\noalign{\smallskip}\hline\noalign{\smallskip}
Random & 2018 Aug 05 & 10--2000 Hz, 8.2 g (rms), 2 mins \\
Sinusoidal  & 2018 Aug 05 & 0--100 Hz, 1.5 g, 2.5 mins \\
Shock & 2018 Aug 09 & 1000 g, twice in each direction \\
Thermal & 2017 July 24-29 & $-$15 to +45 $^\circ$C, 12.5 cycles \\
Thermal-vacuum  & 2018 Sep 18-22 & $-$5 to +30 $^\circ$C, 3.5 cycles, $< 10^{-3}$~Pa \\
\noalign{\smallskip}\hline
\end{tabular}
\end{table*}

We conducted most of the qualification tests for space environment before the payload was delivered to the satellite company for integration, including the mechanical and thermal tests. Due to a tight schedule, the thermal-vacuum test was not done alone, but along with the whole CubeSat after integration. In the thermal-vacuum chamber, an $^{55}$Fe source was used to monitor the detector performance and the results are as expected. Right after each mechanical test, a resonance search was conducted to detect whether or not there was a frequency shift due to mechanical deformation. The test conditions are summarized in Table~\ref{tab:environ}. 

\section{Detector calibration and performance}
\label{sec:cal}

\begin{figure}
\centering
\includegraphics[width=\columnwidth]{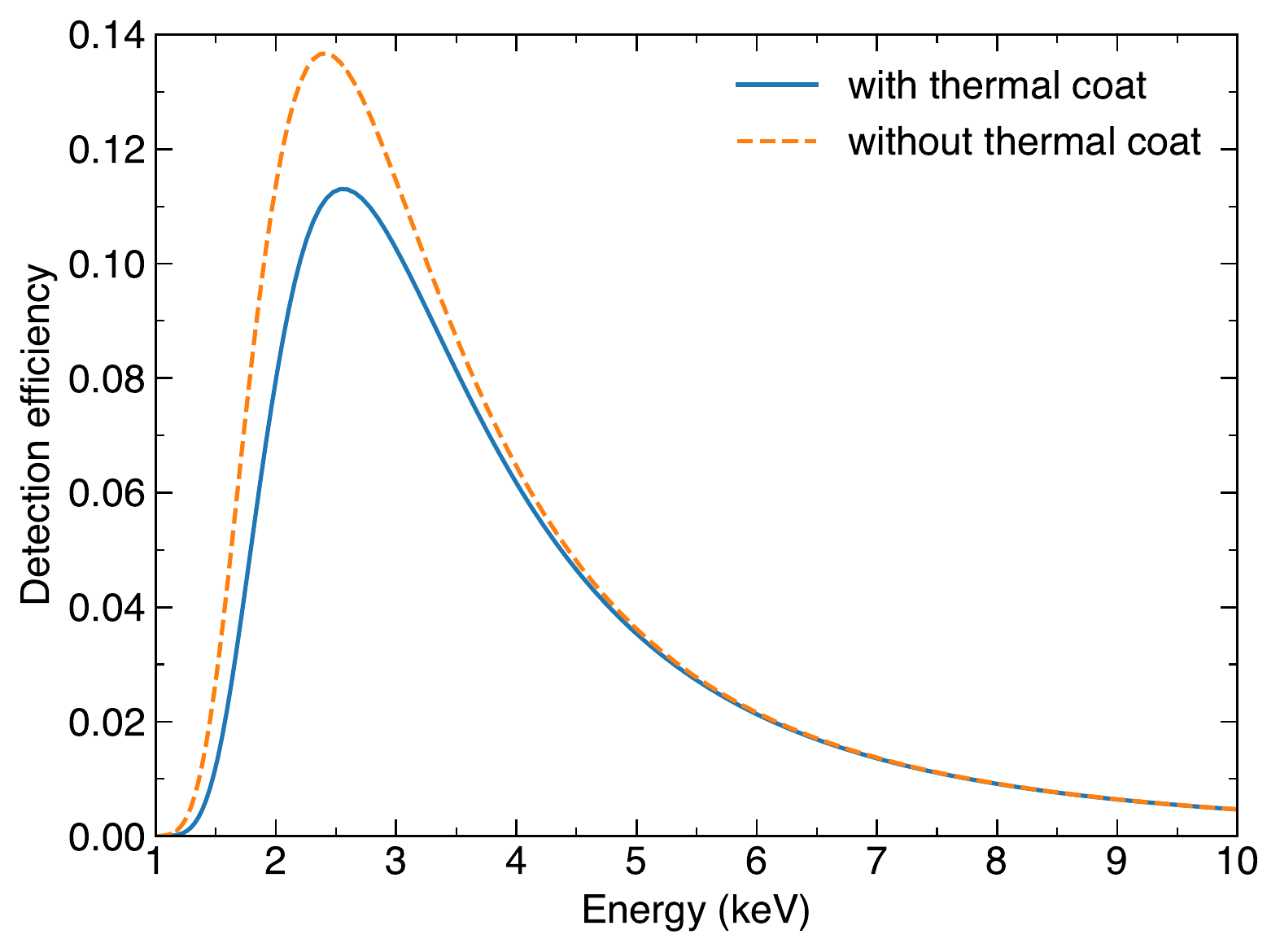}
\caption{Detection efficiency of PolarLight. The dashed line represents the efficiency for the GPD, while the solid line is the efficiency after the thermal coat is covered. }
\label{fig:eff}
\end{figure}

The detection efficiency of the detector is determined by the thermal coat (6 $\mu$m mylar), the beryllium window (100 $\mu$m), and the working gas (1 cm thick DME at a pressure of 0.8 atm).  The thermal coat is optional, but we decided to put it on in order for a stable temperature control. A calculation of the detection efficiency is shown in Figure~\ref{fig:eff}.

\begin{figure}
\centering
\includegraphics[width=\columnwidth]{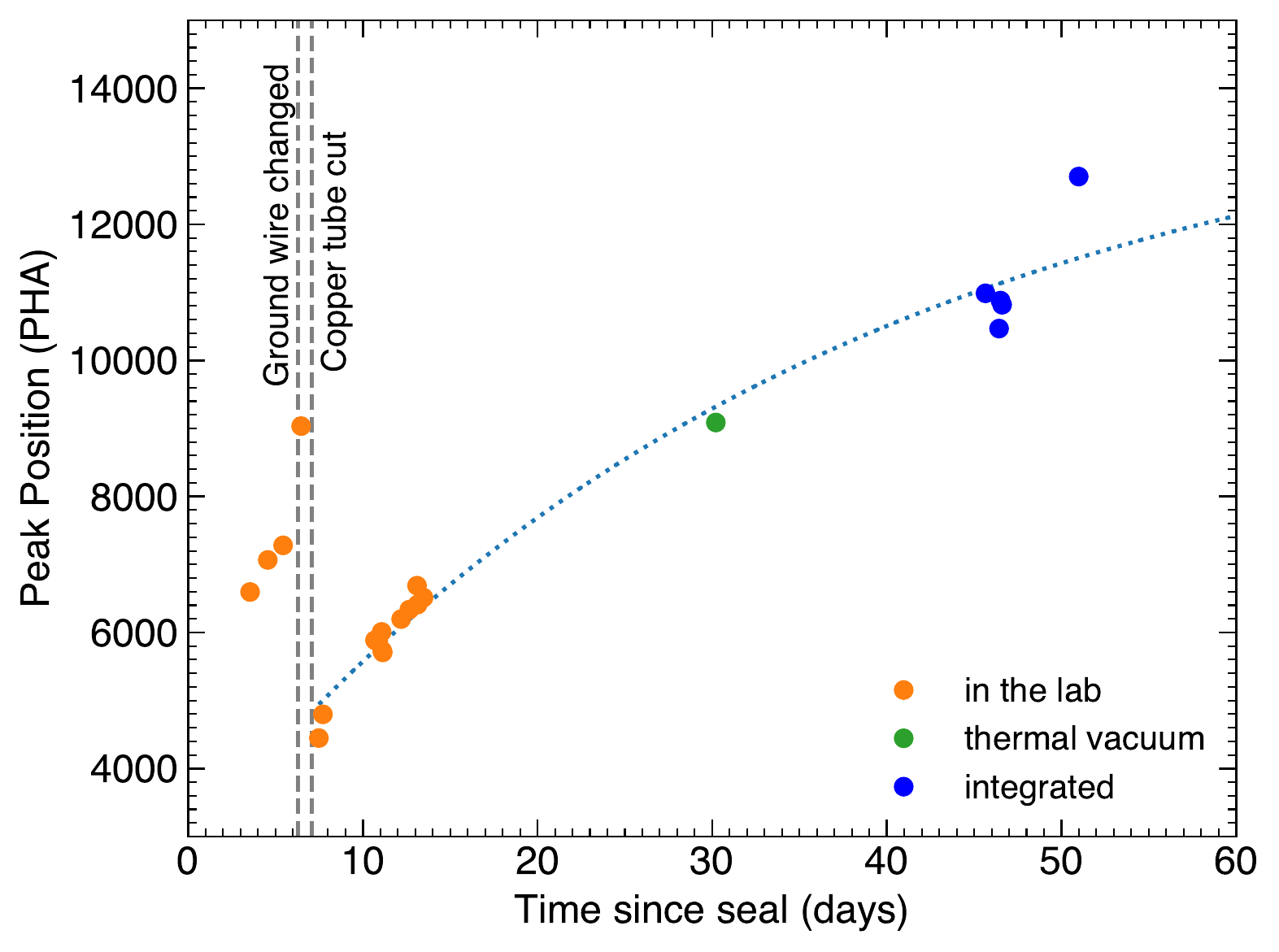}
\caption{Time variation of the detector gain since the seal of the GPD on August 20th, 2018. The dashed lines indicate the times when the ground wire was changed and the copper tube was cut, respectively. The payload was first tested alone in the lab, and then integrated into the CubeSat. }
\label{fig:gain_curve}
\end{figure}

\begin{figure}
\centering
\includegraphics[width=0.8\columnwidth]{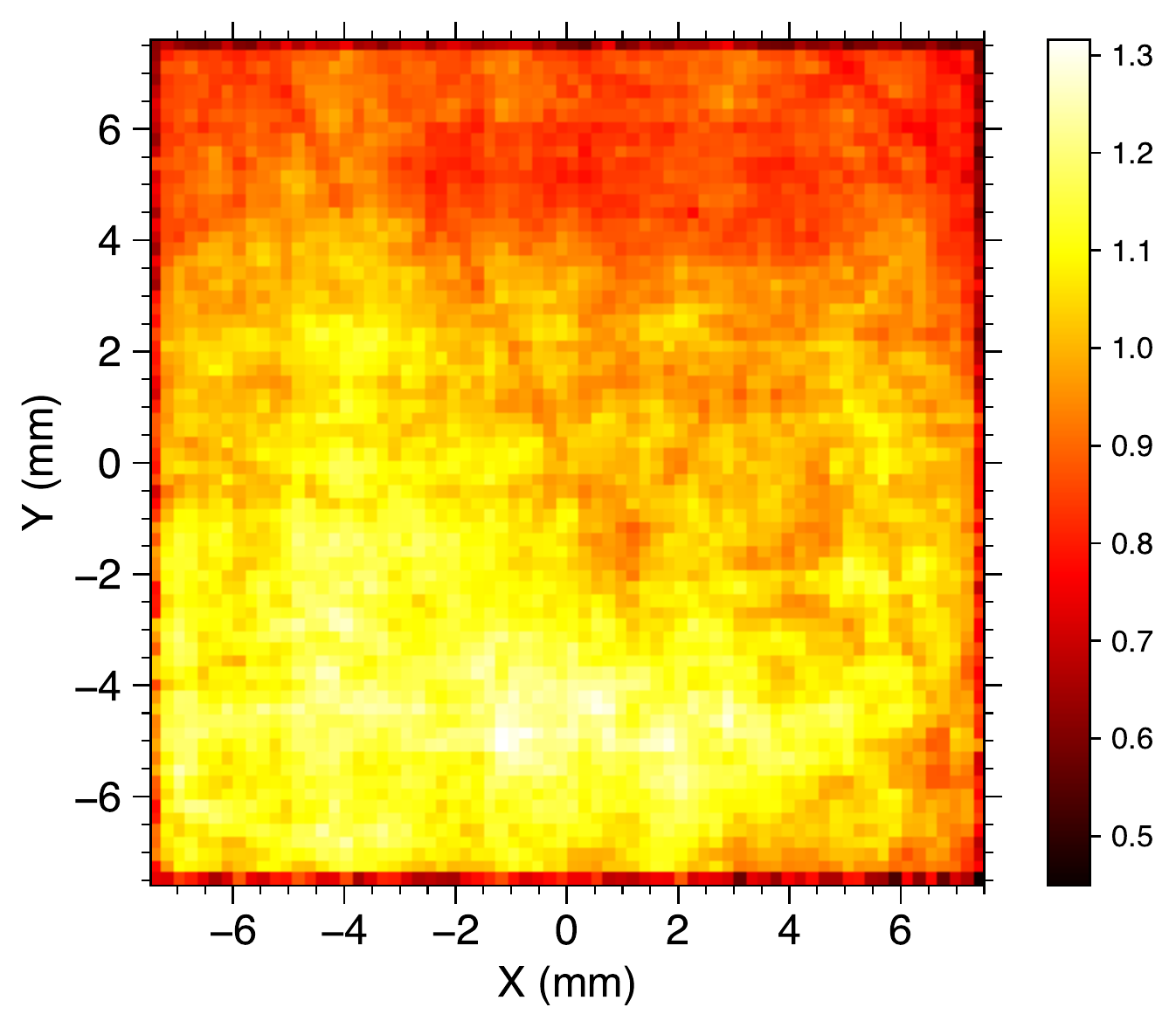}\\
\includegraphics[width=0.8\columnwidth]{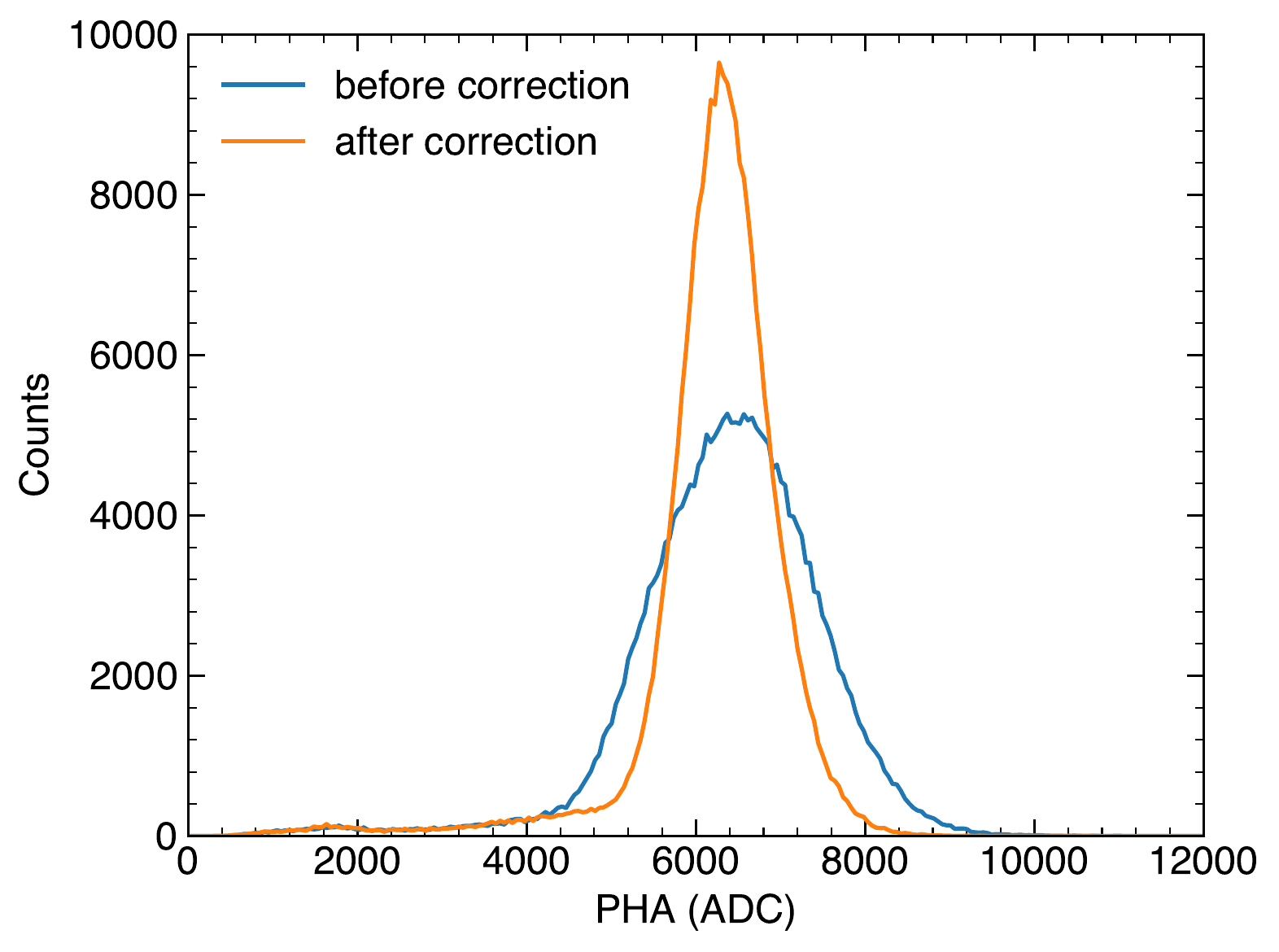}
\caption{\textbf{Top}: gain map ($70 \times 70$) of the GPD across the sensitive surface, normalized to the mean peak position measured with the $^{55}$Fe source. A value larger than unity means the gain in that area is higher than the mean, and vice versa. Pixels around the edge have a low gain because of the loss of charges.  \textbf{Bottom}: $^{55}$Fe spectra before and after correction with the gain map. The FWHM to mean ratio changes from 33.5\% to 18.6\% after the gain map correction.}
\label{fig:gainmap}
\end{figure}

The detector was filled with the working gas and sealed on August 20th, 2018. The gain was found to increase rapidly with time. This is due to the fact that the chamber is sufficiently pumped and the materials inside are adequately degassed, such that our working gas will be absorbed by the materials, especially by those with a relatively high outgassing rate, until an equilibrium is approached. The gain variation curve is shown in Figure~\ref{fig:gain_curve}, where we have converted the peak position to that of $^{55}$Fe with a HV of 3200~V if the measurement was not done in that case.  The electric interface between the payload to the CubeSat is a 21-pin connector, for both power and communication. About 6 days after the detector seal, we realized that a single wire for the ground was insufficient to damp the power surge caused by the HV module, and the HV ground was pulled higher when the HV was on. Thus, we added a second ground wire to solve the problem. This also lead to an increase of the gain as the HV ground was stabilized at zero.  About 7 days after the seal, the detector worked stably and we cut the copper tube, which was used to pump the chamber and fill the gas, to its minimum size. That action may have compressed the gas in the chamber so that the gain dropped (the gain is inversely scaled with the pressure in our case).  Then, laboratory tests and calibrations were conducted before the payload was shipped to the satellite company for integration around 15 days after seal.  Since then, we were not able to test the payload with X-rays for a while except on the 30th day when the whole CubeSat was tested for thermal-vacuum qualification. About one week before the CubeSat was shipped to the launch site, we were allowed to test it and measured a few spectra to verify the detector status. The test was done with a laboratory power supply at first. Then, the battery from the satellite was used (the last point in Figure~\ref{fig:gain_curve}). As one can see, the gain increased significantly. This is probably because the low voltages provided by the satellite battery are not as accurate as expected, and the output PHA is affected. The phenomenon can be repeated using a backup system in the lab. This should be improved in the future by adding a DC-DC module in the payload. Fortunately, the gain variation is less than the spectral resolution ($\sim$20\%) and will not be an issue for our purpose. 

Three levels of data products can be created with the pipeline and saved in the FITS event format. In the science events section, the level-0 file contains information about the time and image. Then,  the level-1 file replaces the image with its characteristics, including the summed pulse height amplitude (PHA) in unit of the ADC value, electron directions estimated using both the major axis method and the impact point method \cite{Bellazzini2003}, number of clusters (how many unconnected pixel islands), number of pixels above the threshold (cluster size), eccentricity, centroid, and the impact point of the track.  Before the image is analyzed, we have adopted an ADC cut of 5 as the threshold (pixels with ADC value $\le$ 5 are set to zero), and applied a median filter on each pixel with its surrounding 7 pixels to remove single, noisy pixels. 

Measured with an $^{55}$Fe source, we found that the gain is not uniform across the detector surface. This is due to the non-uniformity of the GEM thickness~\cite{Takeuchi2014}. We divided the sensitive area into $70 \times 70$ cells, each with a size of $0.214{\rm mm} \times 0.217{\rm mm}$. We collected about 300,000 events with the  $^{55}$Fe source and measured the peak position in each cell to reflect the gain variation.  From here on, we select events that have only one cluster and the cluster size is at least 45 pixels for analysis. A gain map, normalized to the mean peak position of all cells, is created and shown in Figure~\ref{fig:gainmap}. If the gain map is applied on the $^{55}$Fe spectrum, the FWHM to mean ratio changes from 33.5\% to 18.5\%, suggestive of a successful correction.  Then, the level-2 file adds the gain map corrected PHA for each event. 

\begin{figure}
\centering
\includegraphics[width=\columnwidth]{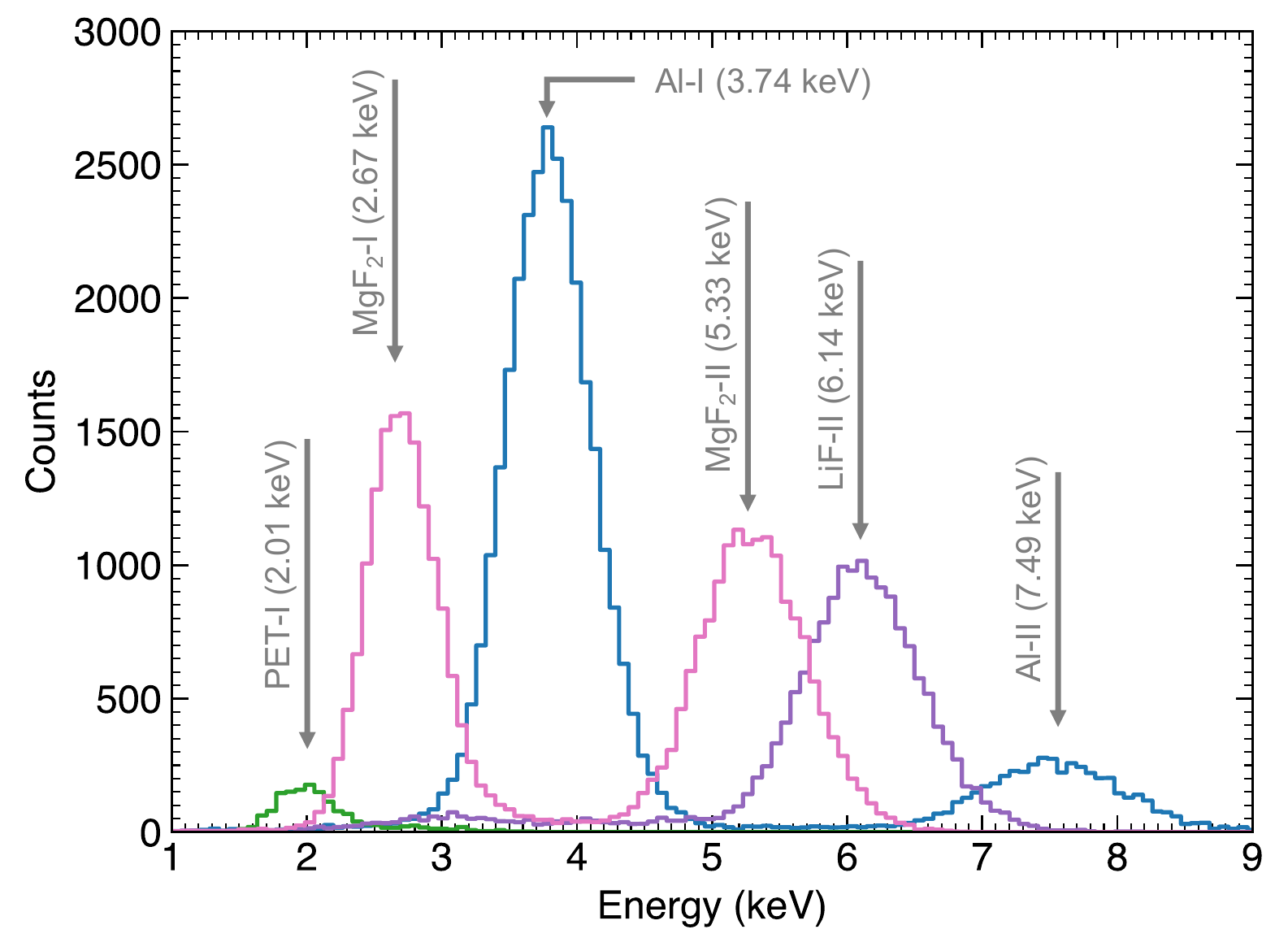}
\caption{Energy spectra measured with 45-degree Bragg diffractions.}
\label{fig:spec}
\end{figure}

We use four Bragg crystals and their 45-degree diffractions to measure the energy spectra and modulation factors of the detector.  A silicon PIN detector was used to take the diffracted energy spectra to check whether or not the peaks appear at the energies as expected.  The measured energy spectra with the PolarLight are shown in Figure~\ref{fig:spec}.  The diffraction energies, along with the measured energy resolutions are listed in Table~\ref{tab:bragg}. The fractional energy resolution (FWHM/$E$) follows an $E^{-1/2}$ relation except at the lowest energy, where the energy resolution is smaller than expected, but this may be due to the low statistics and large error associated with the PET measurement.

\begin{table}
\caption{Bragg crystals, diffracted energies at 45 degrees, and the measured energy resolutions and modulation factors.}
\label{tab:bragg}
\begin{tabular}{lcccc}
\hline\noalign{\smallskip}
Crystal & Order & $E$  & FWHM/$E$ & $\mu$ \\
&&(keV)&&\\
\noalign{\smallskip}\hline\noalign{\smallskip}
PET & I & 2.01 & $0.230 \pm 0.008$ & $0.136 \pm 0.047$ \\
MgF$_2$ & I & 2.67 & $0.230 \pm 0.002$ & $0.211 \pm 0.015$ \\
Al & I & 3.74 & $0.201 \pm 0.001$ & $0.420 \pm 0.009$ \\
MgF$_2$ & II & 5.33 & $0.175 \pm 0.002$ & $0.513 \pm 0.012$ \\
LiF & II & 6.14 & $0.164 \pm 0.001$ & $0.568 \pm 0.010$ \\
Al & II & 7.49 & $0.165 \pm 0.006$&  $0.665 \pm 0.022$ \\
\noalign{\smallskip}\hline
\end{tabular}
\end{table}

\begin{figure}
\centering
\includegraphics[width=0.49\columnwidth]{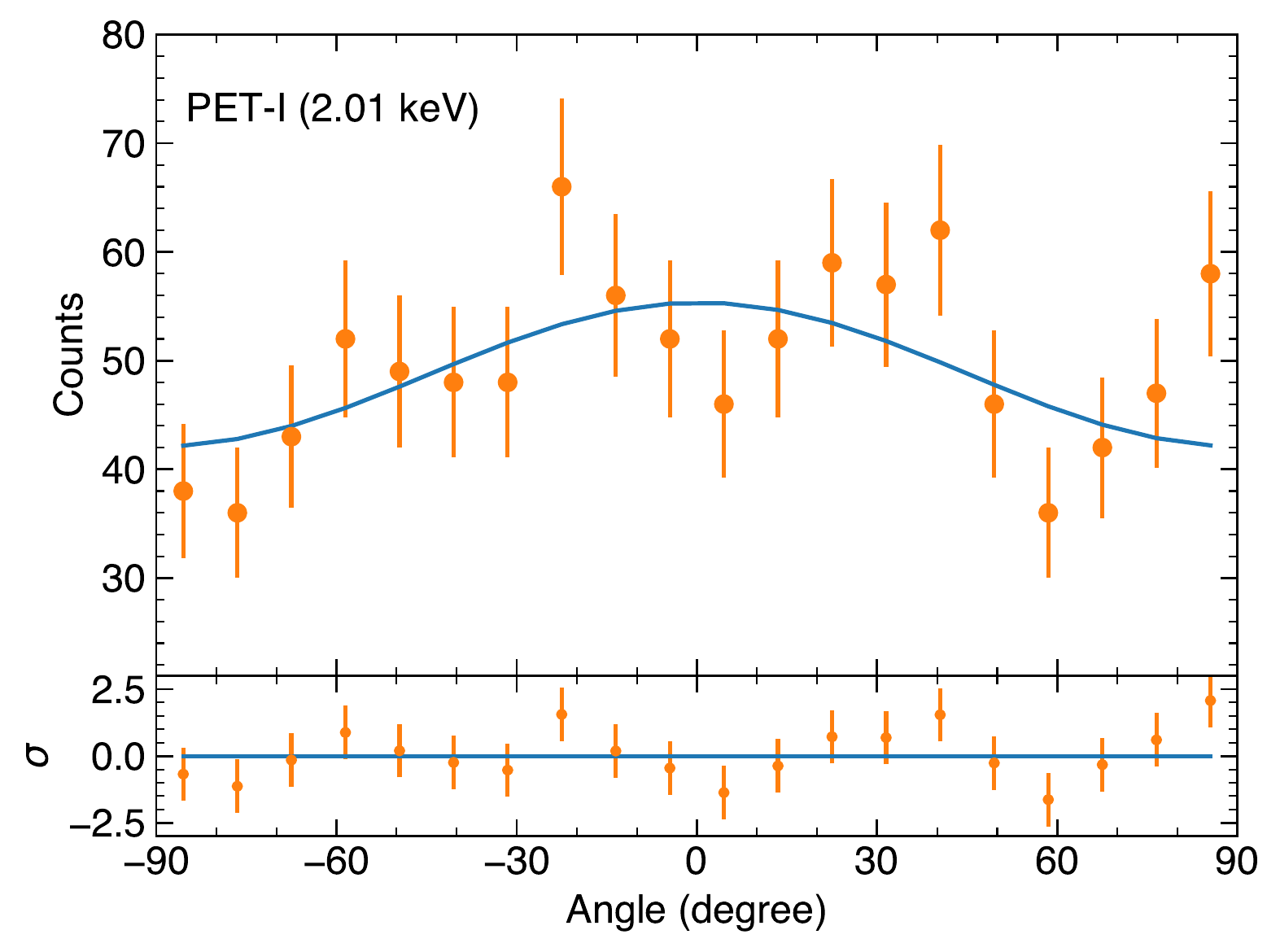}
\includegraphics[width=0.49\columnwidth]{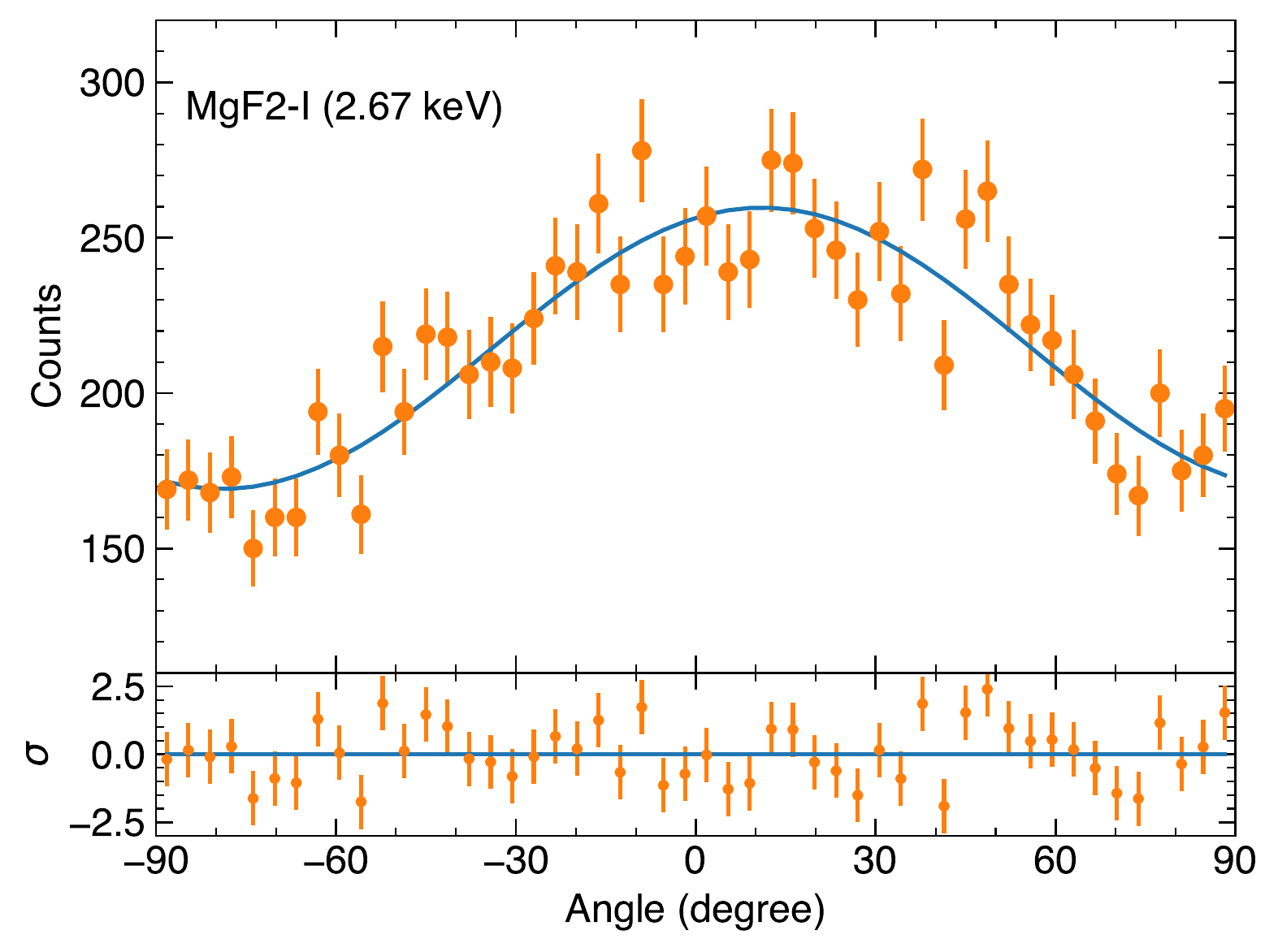}\\
\includegraphics[width=0.49\columnwidth]{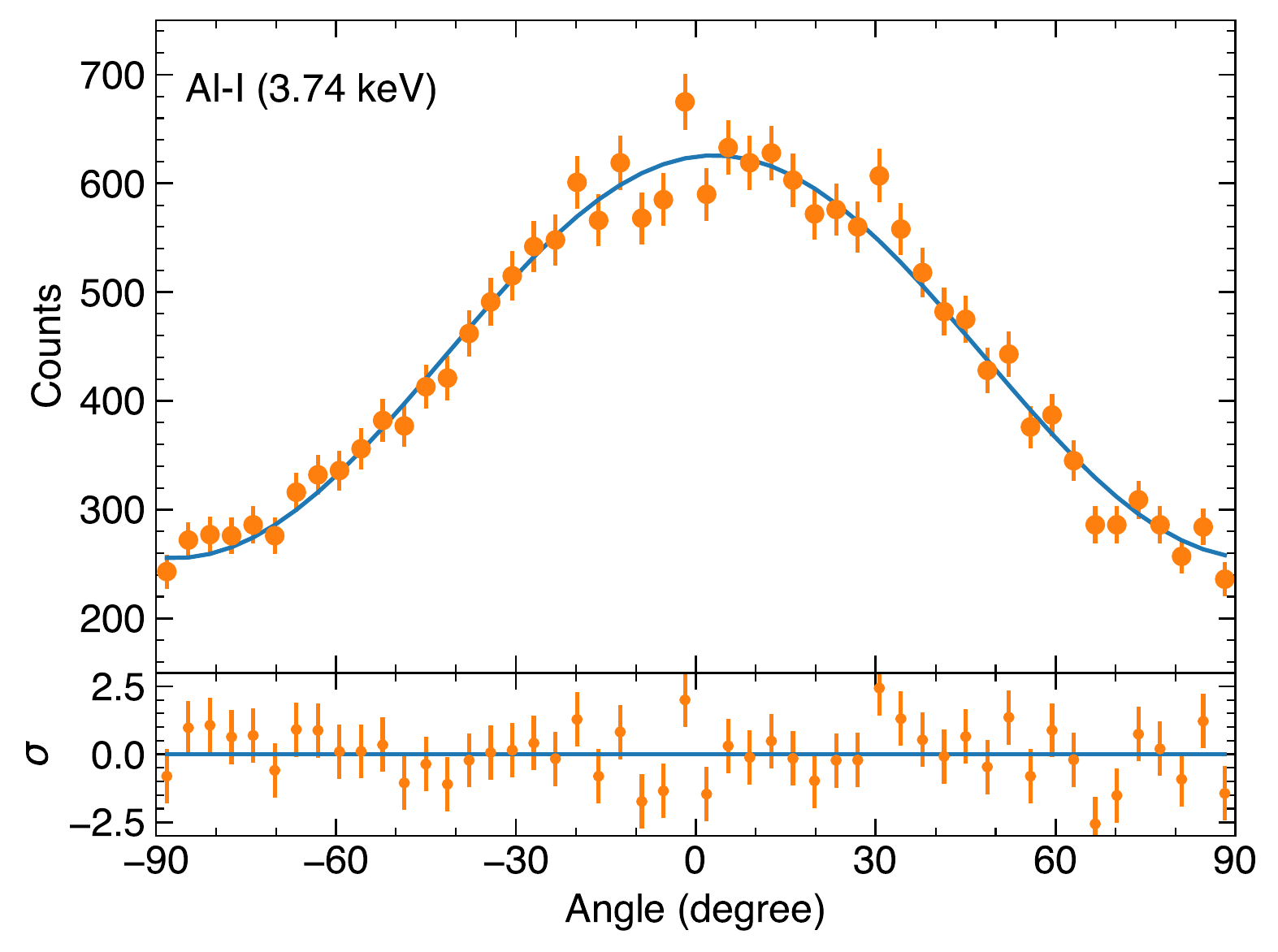}
\includegraphics[width=0.49\columnwidth]{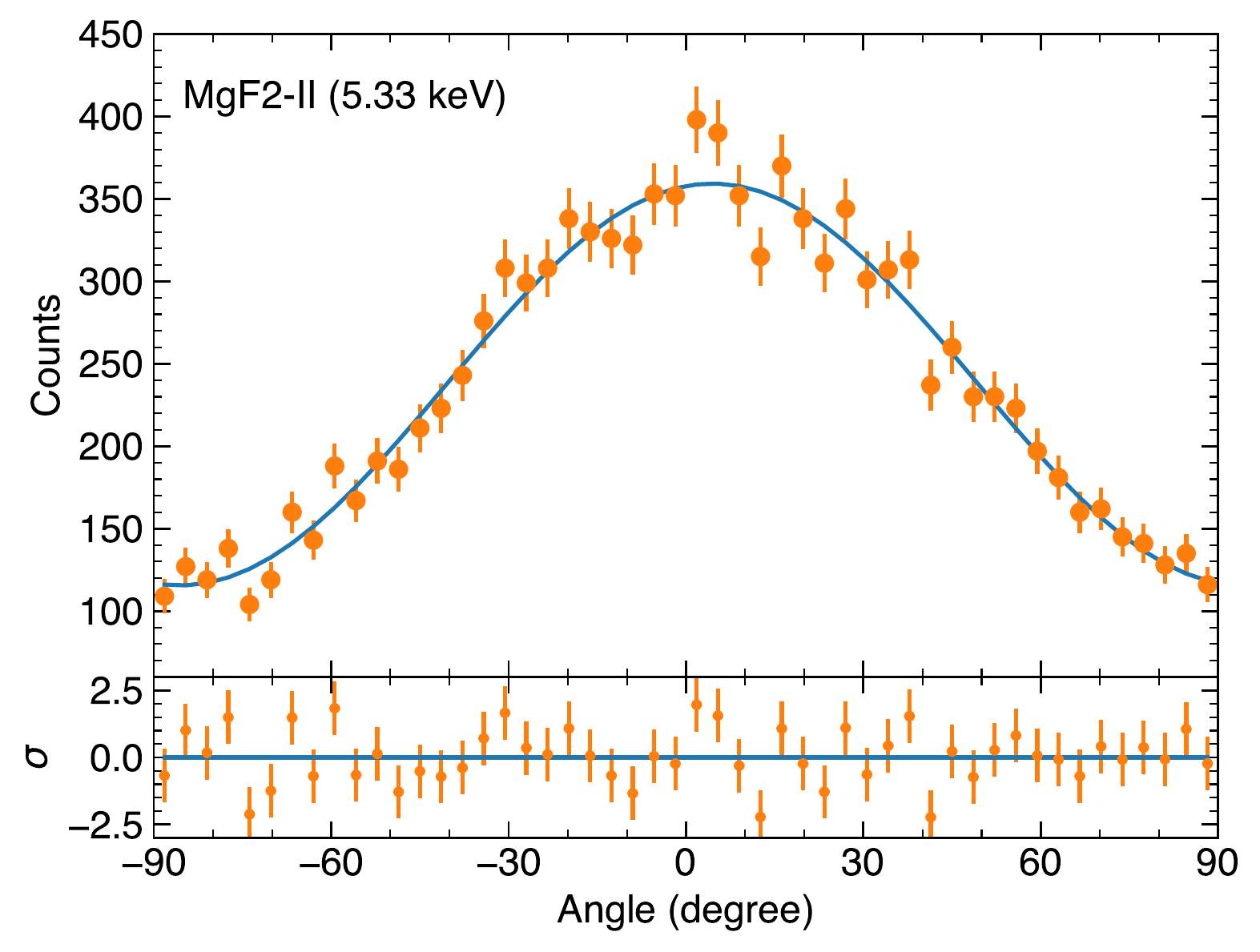}\\
\includegraphics[width=0.49\columnwidth]{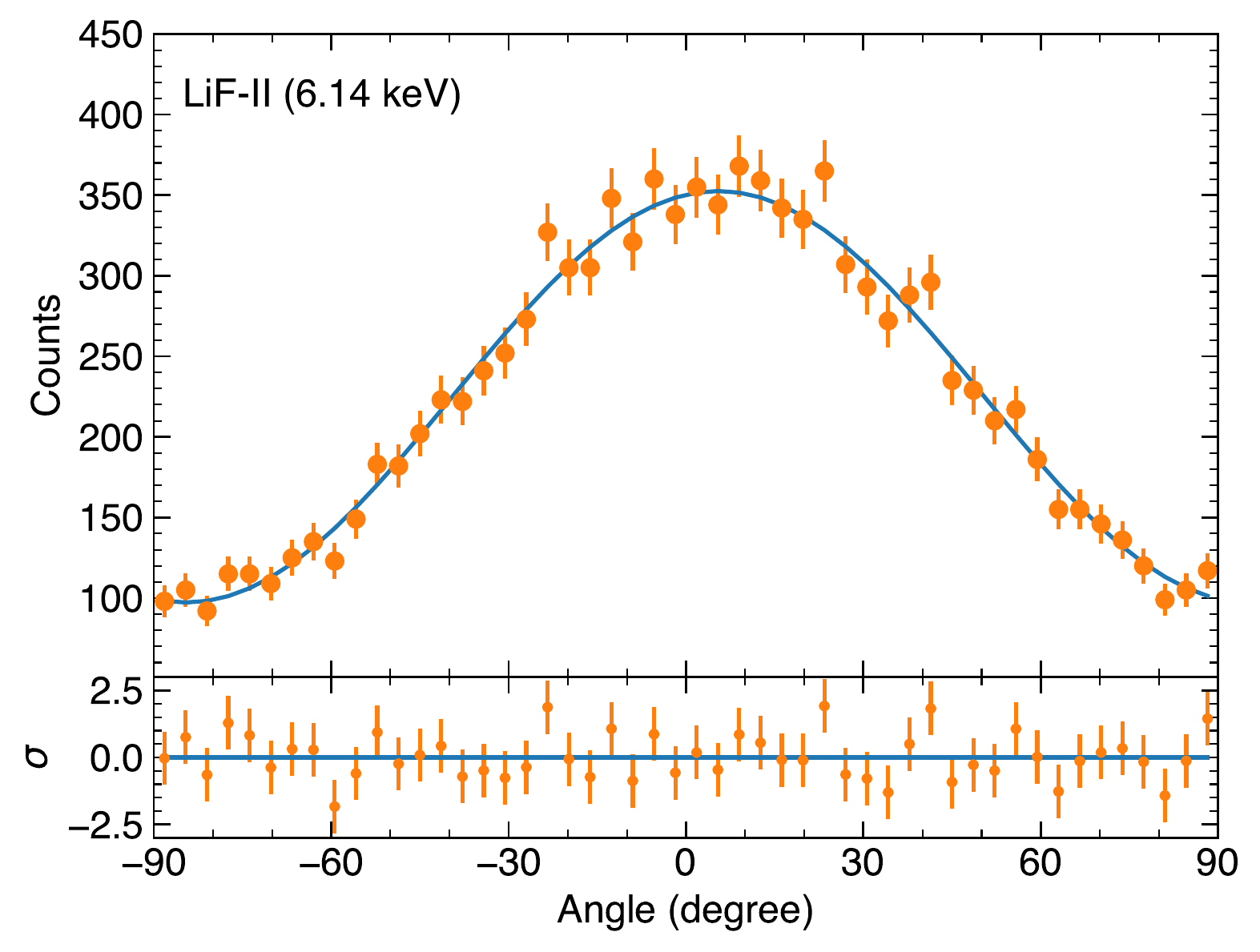}
\includegraphics[width=0.49\columnwidth]{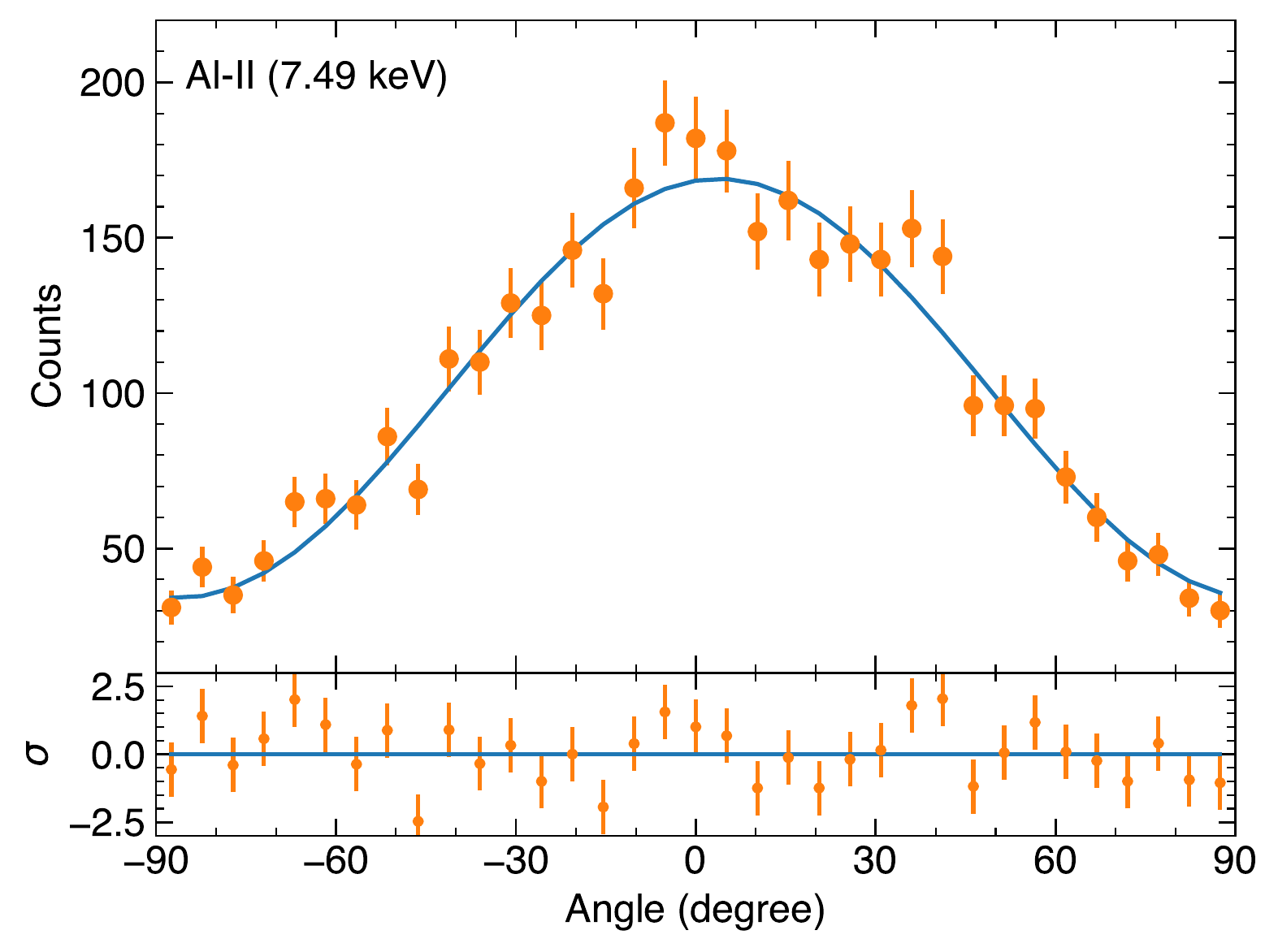}
\caption{Modulation curves measured with 45-degree Bragg diffractions.}
\label{fig:modcuv}
\end{figure}

\begin{figure}[h!]
\centering
\includegraphics[width=0.8\columnwidth]{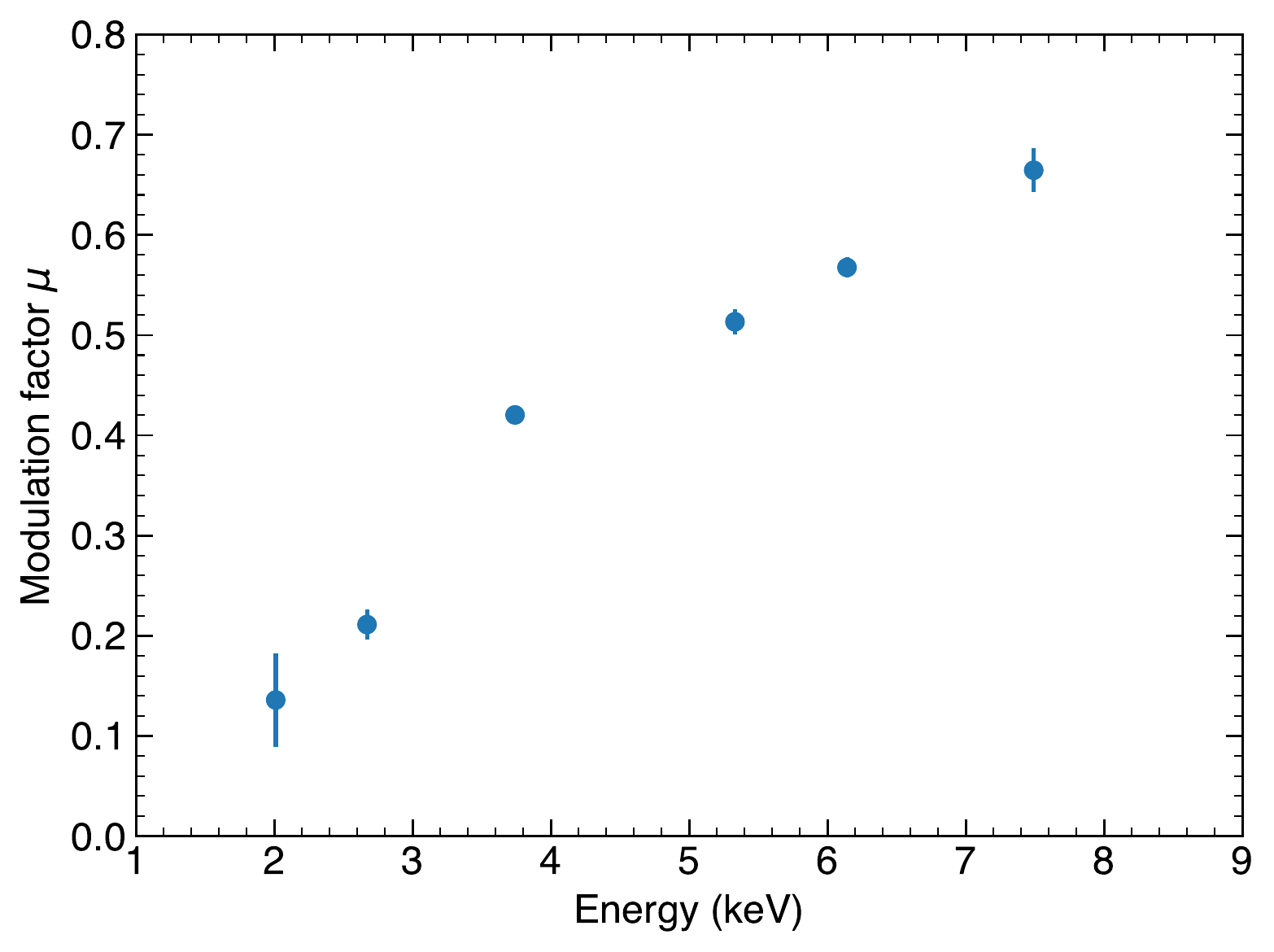}\\
\includegraphics[width=0.8\columnwidth]{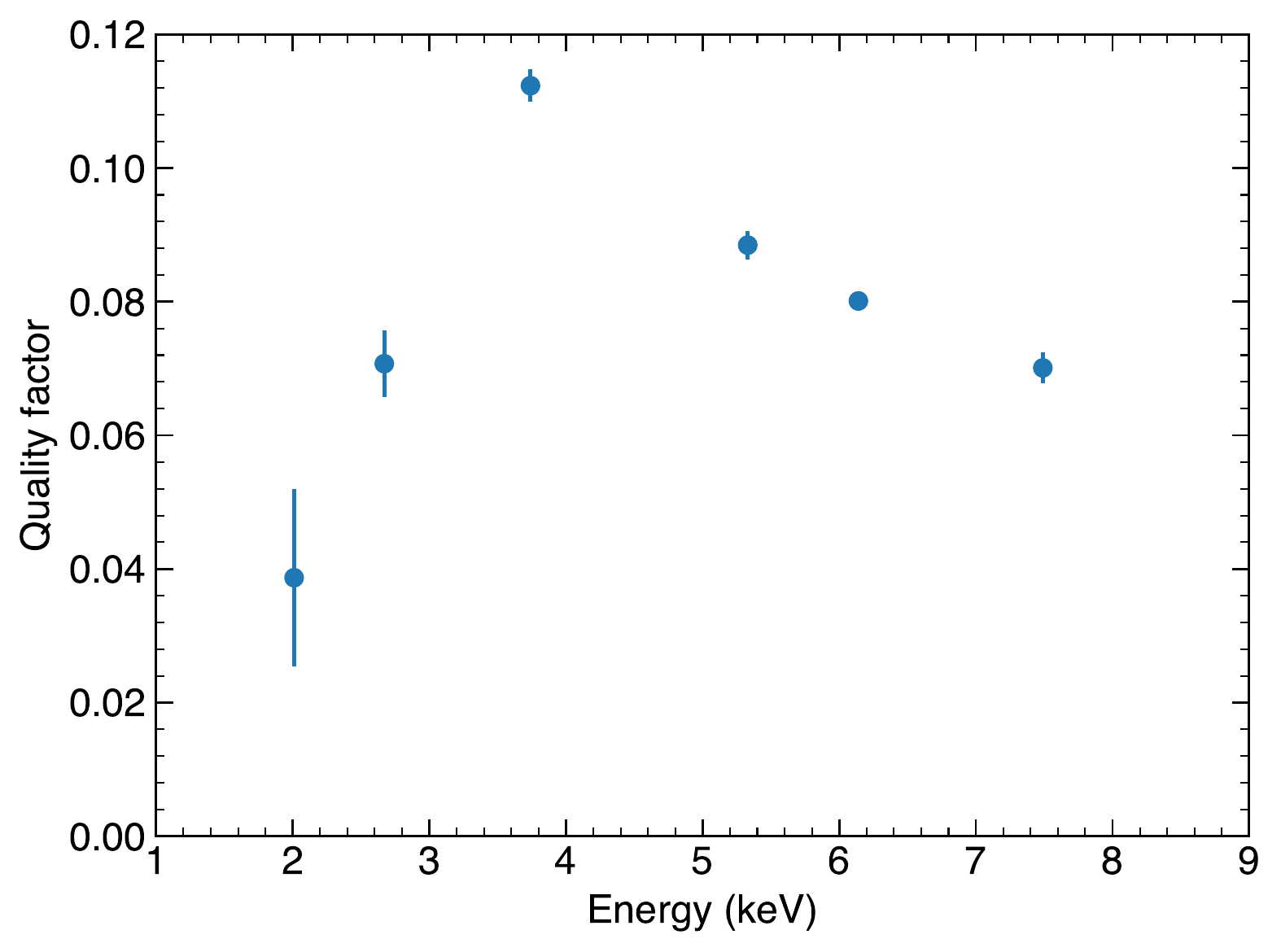}
\caption{Modulation factor (\textbf{top}) and the quality factor (\textbf{bottom}) for PolarLight. The quality factor is defined as modulation factor times the square root of the efficiency.}
\label{fig:mod}
\end{figure}

As 45-degree diffraction produces a fully polarized X-ray beam, the same data are used to calculate the modulation factors. Following the literature \cite{Muleri2010,Li2015}, we discarded 25\% of the events with the lowest eccentricity.  The emission angle distributions of electrons are shown in Figure~\ref{fig:modcuv}, and fitted with a modulation function, $A + B\cos^2(\phi - \phi_0)$. At energies below 3~keV, the major axis direction is adopted as the emission angle, while above 3~keV, the impact point method is used. The degree of modulation,  $(\max - \min) / (\max + \min)$, is then calculated and displayed in Table~\ref{tab:bragg} and Figure~\ref{fig:mod} (top panel), which is the modulation factor ($\mu$; degree of modulation resulted from fully polarized X-rays) of the instrument.  As the diffracted beam cannot illuminate the whole detector plane at a close distance, we thus checked the positional uniformity at 9 ($3 \times 3$) points on the surface of the detector and found that the modulation factor and position angle do not show a detectable change with respect to the location. 

The sensitivity of an X-ray polarimeter is proportional to the quality factor, which is a product of the modulation factor and the square root of the detection efficiency.  In Figure~\ref{fig:mod} (bottom panel), we display the quality factor as a function of energy. As one can see, the sensitivity of the PolarLight peaks around 4~keV.

We note that our results are well consistent with those reported by previous studies, except at 2.7~keV for the MgF$_2$-I line, where a modulation factor of 0.27 was reported \cite{Muleri2010,Muleri2012,Li2015}. The major difference between our setup and the previous is the GEM pitch. For PolarLight, the GEM pitch is 100~$\mu$m and larger than before. For X-rays of higher energies, the electron track is long and the result may not be limited by the pitch, while for the lowest energy (2~keV), the track is unresolved so that the pitch is not important. Around 3~keV, the pitch may have played the most important role in sampling the electron track. This may explain why consistent modulation factors can be obtained at other energies.  More experiments and simulations should be done in the future to investigate this problem. 

\section{Sensitivity and targets}

As the boarder region may suffer from charge loss (see the gain map in Figure~\ref{fig:gainmap}), we only extract the $\pm 7$ mm region around the center for science analysis. With the detection efficiency and open fraction of the collimator quoted above, and taking into account the source spectrum and flux, the count rate expected from the brightest X-ray sources in several energy bands are listed in Table~\ref{tab:rate}, along with the references from which the source spectra are adopted. The 2--8 keV band is quoted as the full band of the detector, as copper K$\alpha$ line originated from the GEM foil appears above this band. The 3--5 keV band is the energy range where the sensitivity is the highest, and the 4--8 keV band has the highest modulation factor. The average modulation factor is 0.25, 0.37, and 0.49, respectively in the three bands, weighted using a detected Crab spectrum. The choice of the energy band relies on the specific scientific objective.

\begin{figure}
\centering
\includegraphics[width=\columnwidth]{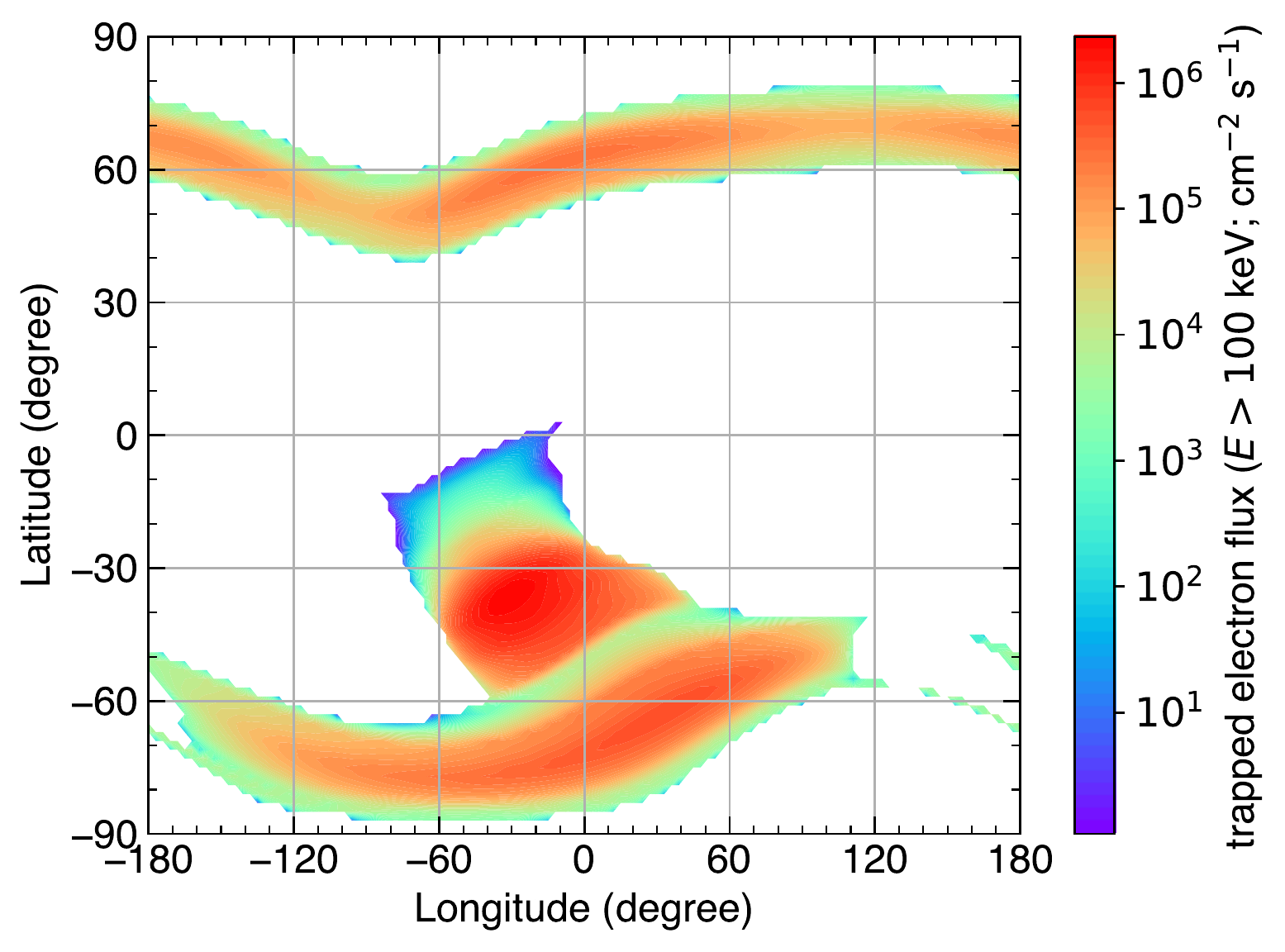}
\caption{Flux map for trapped electrons in the PolarLight orbit with energies above 100~keV (those that can penetrate the beryllium window). The data are obtained from SPVINS (www.spenvis.oma.be) using the AE-8 model at the solar minimum.}
\label{fig:emap}
\end{figure}

The background rate due to the CXB is found to be a few times $10^{-4}$~\cts\ and is negligible compared with the expected flux from the brightest X-ray sources. However, as the CubeSat is in a polar orbit, where the particle flux is high near the polar and the south Atlantic anomaly (SAA) regions (see Figure~\ref{fig:emap}), the particle induced background, especially the delayed background due to activation in the high flux region, may be dominant. This is still unknown at the point of submission.  

The minimum detectable polarization (MDP) at a confidence level of 99\% is usually quoted as the sensitivity of a polarimeter \cite{Weisskopf2010,Strohmayer2013}, 
\begin{equation}
\label{mdp}
    {\rm MDP_{99}} = \frac{4.29}{\mu S} \sqrt{\frac{S + B}{T}},
\end{equation}
where $S$ is the source count rate, $B$ is the background count rate, $\mu$ the modulation factor, and $T$ is the total exposure time.  For the Crab nebula, which has a known degree of polarization of 19\%, its polarization signal in the 2--8 keV band can be detected with an exposure time of a few times $10^5$~s if the background rate is $\sim$1~\cts. If the background exceeds 5~\cts, the polarization from Crab is no longer detectable even with a net exposure of $10^6$~s. Because the HV cannot be powered on in regions of high particle flux,  also due to Earth occultation, the effective observing time is less than one half of the total operation time. 

\begin{table*}
\caption{Expected count rates for bright X-ray sources in different energy bands measured with PolarLight.}
\label{tab:rate}
\centering
\begin{tabular}{lcccc}
\hline\noalign{\smallskip}
 & \multicolumn{3}{c}{Rate (\cts)}  & Reference \\
  \noalign{\smallskip}\cline{2-4}\noalign{\smallskip}
 & (2--8 keV) & (3--5 keV)  & (4--8 keV) &  \\
\noalign{\smallskip}\hline\noalign{\smallskip}
Crab & 0.20 & 0.068 & 0.029 & \cite{Kirsch2005}  \\
GRS 1915+105 (thermal state) & 0.16 & 0.078 & 0.040 & \cite{McClintock2006}  \\
Cygnus X-1 (low/hard state) & 0.078 & 0.029 & 0.014 & \cite{Sugimoto2016}  \\
Scorpius X-1  & 3.5 & 1.2 & 0.46 & \cite{Church2012} \\
\noalign{\smallskip}\hline
\end{tabular}
\end{table*}

\begin{figure}
\centering
\includegraphics[width=\columnwidth]{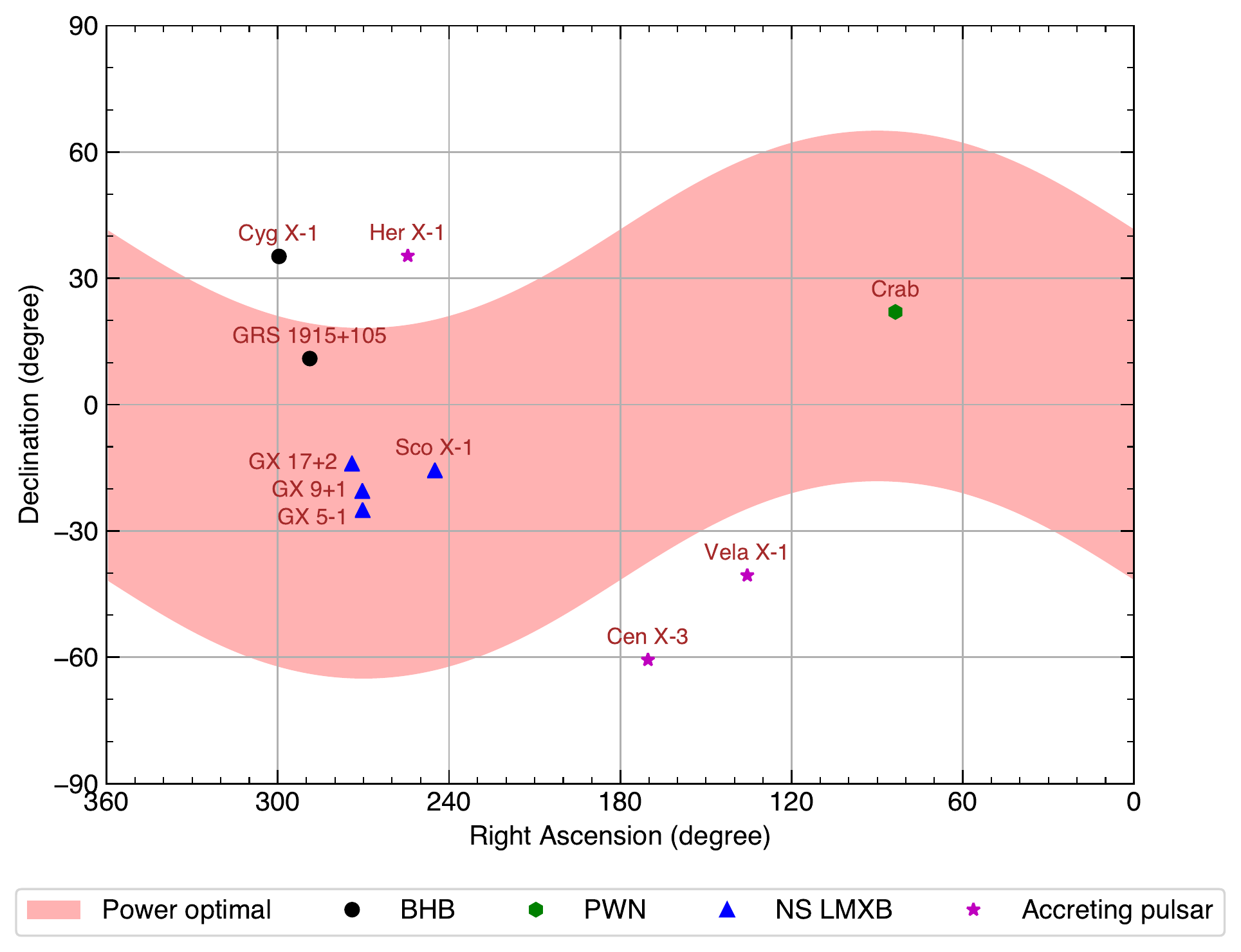}
\caption{Possible targets for PolarLight, including black hole binaries (BHBs), pulsar wind nebulae (PWNe), neutron star low-mass X-ray binaries (NS LMXBs), and accreting pulsars (highly magnetized systems). If the sources are in the shaded region, they may be pointed by PolarLight continuously without a power issue.  For sources outside the shaded region, they can be observed but the observation has to be interrupted due to battery charge and/or Earth occultation due to a large angle between the Sun and solar panel. }
\label{fig:obs}
\end{figure}

Anyway, bright X-ray sources on the sky, including pulsar wind nebulae and accreting compact objects, could be targets of PolarLight (Figure~\ref{fig:obs}).  We note that strong magnetic systems, such as accreting pulsars, are of particular interest because of their potentially high degree of polarization. Limited by the power, some regions (shaded in Figure~\ref{fig:obs}) in the sky may be pointed by the PolarLight continuously. While other regions, due to large angles between the solar panel and the Sun, the observation can be conducted but has to be intermittent due to battery charge and/or Earth occultation. 

\section{Discussion and conclusion}

Here, we report on the design and ground test results for PolarLight, which did the first flight test for the GPD polarimeter. The main purpose for PolarLight is to demonstrate the technique in space and reveal potential issues with the detector design, which will be valuable for future missions like eXTP as the same detector will be used. Limited by the tight schedule and constraints on resources, some issues already emerged during the ground test and calibration.  We list the lessons gained so far:

\begin{itemize}

  \item A DC-DC module is necessary to provide a stable power supply for the detector. 
  
  \item The detector needs be sealed at least a few months in advance, so that the gain will be stable at the time of launch.  
  
  \item The gain is sensitive to the gas pressure. The equilibrium between the absorption and outgassing inside the chamber is a function of temperature.  Change of the storage temperature may lead to a change of the gain temporarily. Once launched, the detector temperature would best be controlled in a narrow range, no matter in operation or not. 
  
  \item A gain calibration seems useful every time when an observation is done. Online calibration is impossible for a CubeSat, but has been designed for future large missions. 
  
  \item Whether or not the 100-$\mu$m-pitch GEM can produce a modulation factor as high as that with a 50-$\mu$m-pitch GEM needs in-depth investigations. A coarse pitch allows for a thicker GEM (100 $\mu$m) and a larger gain. 
 
 \end{itemize}

After the CubeSat finishes the communication and attitude test, we will start the full test of the PolarLight and investigate the in-orbit background and its influence to the polarization measurement. A new flight is planned in 2019 with an improved design. A 50 $\mu$m window will be used and known issues will be fixed.  

\begin{acknowledgements}
We thank Wenfei Yu for helpful discussions about the choice of targets, and Spacety for helping with the test. HF acknowledges funding support from the National Natural Science Foundation of China under the grant Nos.\ 11633003 and 11821303, and the National Key R\&D Program of China (grant Nos.\ 2018YFA0404502 and 2016YFA040080X). The initial development of the GPD concept was funded by the Italian National Institute for Nuclear Physics (INFN), the Italian Space Agency (ASI), and the Italian National Institute for Astrophysics (INAF).
\end{acknowledgements}


\end{document}